\documentclass[prb,twocolumn,amsmath,amssymb,floatfix,footinbib,superscriptaddress,bibnotes]{revtex4-2}
\usepackage{graphicx}
\usepackage{dcolumn}
\usepackage{bm}
\usepackage{bbm}
\usepackage{diagbox}
\usepackage{braket}
\usepackage{titlesec}
\usepackage{enumerate}
\usepackage{subfigure}
\usepackage{epstopdf}
\usepackage[colorlinks,linkcolor=red,anchorcolor=black,citecolor=blue]{hyperref}
\usepackage{amssymb}
\usepackage{framed}
\usepackage{tabularx}
\usepackage{booktabs}
\usepackage{float}
\usepackage[utf8]{inputenc}
\usepackage{amsfonts,mathtools,physics,url,slashed}
\usepackage{caption}
\usepackage{breqn}
\newcommand{\avg}[1]{\langle#1\rangle}
\newcommand{\bigavg}[1]{\biggr\langle#1\biggr\rangle}

\newcommand{\sign}{\text{sgn}}

\begin{document}
\title{Non-Fermi Liquids from Dipolar Symmetry Breaking}
\author{Amogh Anakru}
\email{apa6106@psu.edu}
\author{Zhen Bi}
\email{zjb5184@psu.edu}
\affiliation{Department of Physics$,$ The Pennsylvania State University$,$ University Park$,$ Pennsylvania$,$ 16802$,$ USA}

\begin{abstract}
The emergence of fractonic topological phases and novel universality classes for quantum dynamics highlights the importance of dipolar symmetry in condensed matter systems. In this work, we study the properties of symmetry-breaking phases of the dipolar symmetries in fermionic models in various spatial dimensions. In such systems, fermions obtain energy dispersion through dipole condensation. Due to the nontrivial commutation between the translation symmetry and dipolar symmetry, the Goldstone modes of the dipolar condensate are strongly coupled to the dispersive fermions and naturally give rise to non-Fermi liquids at low energies. The IR description of the dipolar symmetry-breaking phase is analogous to the well-known theory of a Fermi surface coupled to an emergent U(1) gauge field. We also discuss the crossover behavior when the dipolar symmetry is slightly broken and the cases with anisotropic dipolar conservation. 
\end{abstract}
\maketitle

\section{Introduction}

Symmetry is a crucial organizing principle in condensed matter physics. Recently, there has been a significant effort in quantum condensed matter physics to study various generalized forms of symmetries\cite{mcgreevy_generalized_symmetries}. These generalized symmetries are more closely related to underlying geometric structures compared to ordinary global symmetry. The profound consequences of these generalized symmetries are not fully understood yet. One important family of generalized symmetry is multipolar symmetry\cite{Gromov_Multipole}. A system has multipolar symmetry if its Hamiltonian has a conserved charge and conserves certain multipolar moments associated with the said charge. Multipolar symmetry has a unique mathematical structure, leading to various intriguing dynamical phenomena. For instance, fracton topological phases of matter\cite{fractonreview1,fractonreview2} can arise due to multipolar symmetry\cite{Gromov_Multipole, bulmash2023multipole}, where charged excitations' dynamical properties are highly constrained. Therefore, multipolar symmetry is a robust pathway to ergodicity breaking and exotic universality classes of quantum dynamics\cite{pai_pretko_nandkishore_fractonic_random_circuits,sala2020_ergodicity_from_dipole_cons,aidelsburger_fragmentation_tilted,fractonhydro1,fractonhydro2,fractonhydro3,subdiffusion1,subdiffusion2,subdiffusion3,freezing,khemani_hermele_nandkishore_tilted}. Dipole conservation may be realized as an approximate conservation law in systems with a linearly varying potential \cite{khemani_hermele_nandkishore_tilted,huse_bakr_tilted_lattice}, and such behavior has been explored in experimental settings with cold atoms in a strongly tilted optical lattice \cite{scherg_aidelsburger_2021observing}.

Spontaneous symmetry breaking is a common phenomenon in many condensed matter systems, including magnets, crystals, and superconductors. The presence of dynamic constraints due to multipolar symmetry makes the properties of the spontaneous breaking of these symmetries different from ordinary global symmetry \cite{PengYe1,PengYe2}. Recently, a set of generalized Mermin-Wagner-type constraints for multipolar symmetries was discussed, showing a clear distinction from ordinary continuous global symmetry \cite{MultipolarMerminWagner,kapustin_dipole}. Lake et al. \cite{LakeDBHM} studies a generalization of the Bose-Hubbard model to systems with dipolar symmetries. Besides the Mott insulating phase, the model displays diverse phases distinguished by dipole condensation or single-boson condensation. The single-boson-condensed phase, influenced by dipolar symmetry, has zero conductance despite being compressible, which sets it apart from a traditional superfluid. Spontaneous symmetry-breaking of dipolar symmetry has been shown to give rise to highly unusual hydrodynamic modes\cite{Lucas_Dipolar_Goldstone_HD,stahl2023fracton,fractonhydro1,fractonhydro2,fractonhydro3}.

This work examines fermionic models that adhere to dipolar symmetry. Specifically, we investigate fermions residing on a $d$-dimensional cubic lattice, subject to the usual $U(1)$ charge conservation symmetry and an additional set of dipolar $U(1)$ symmetries along each spatial direction, enforcing the preservation of dipole moments in each direction. The lattice sites are labeled by $\vec{r}_{\bf{j}}=\sum_a j_a\vec{a}$, where $\vec{a}$ denotes the primitive lattice vectors, and the charge operators for the $d$ conserved dipole moments are denoted by $\hat{Q}_a =\sum_{\bf{j}} \vec{a}\cdot \vec{r}_{\bf{j}}\hat{n}_{\bf{j}}$. Here, $\hat{n}_{\bf{j}}$ is the fermion number operator on site $\bf{j}$. It is evident that the dipolar $U(1)$ symmetries, together with the charge-$U(1)$ and lattice translations, comprise a multipolar algebra with a nontrivial commutation relation between the dipole and the translations \cite{Gromov_Multipole}. This forbids conventional hopping terms $c^\dag_{\bf{j}}c_{\bf{j}+\bf{a}}$ and restricts the permissible quadratic fermion operators to number operators on each site. As a result, any nontrivial Hamiltonian must feature at least quartic terms. 

Although predicting the phase diagram of an interacting dipolar symmetric fermionic model is challenging, our study concentrates on the universal properties of spontaneous dipolar symmetry-breaking phases. Dipole condensation results in a finite single-particle hopping amplitude for fermions, enabling the formation of a mobile Fermi surface, which is the primary focus of this work. Interestingly, such interaction-generated mobility has also been studied in the context of composite Fermi liquids in the half-filled Landau level\cite{HLR}. Although at the mean-field level, one might naively expect a Fermi liquid state at low energies in the dipole condensed phase, we demonstrate that the mobile low-energy fermions are strongly coupled to the Nambu-Goldstone modes (NGBs) arising from the spontaneously broken dipolar symmetries due to the multipolar algebra structure\cite{CriterionVishwanath}. Remarkably, the low-energy effective field theory of the dipole condensed phase is intimately related to that of the spinon fermi surface coupled to a dynamical $U(1)$ gauge field\cite{HLR, dopemott, lohneysen_HMM_theory, senthil_spinon_FS,mross_mcgreevy_liu_senthil_NFL_eps_N, sslee2009,sachdevQPT, metlitski_sachdev_ising,mandal_transverse_gauge_field,sachdev_criticalFS_1, sachdev_criticalFS_2}. Consequently, the low-energy fermions are expected to form a non-Fermi liquid state at generic fillings in 2 spatial dimensions.

The remainder of this paper is organized as follows. In Sec. \ref{Sec:MF}, we present mean-field results for some toy models with dipolar symmetries, aiming to demonstrate the possibilities of spontaneous dipolar symmetry-breaking phases. We then shift our focus to low-energy effective theories of the dipole-condensed phase, where the interaction between the resulting fermi surface and the NGBs can be determined by symmetry considerations alone. In Sec. \ref{Sec:1D}, we examine the dipole condensed phase in 1D systems, where we find that the coupling to fermions alters the Mermin-Wagner criteria for dipolar symmetry-breaking in an intriguing way. We explore the 2D case in Sec. \ref{Sec:2D}, utilizing the analogy of our theory to the well-known problem of a spinon Fermi surface coupled to an emergent gauge field to describe the non-Fermi liquid fixed point. This analysis enables us to predict various universal features of the dipole condensed phase, including the behavior of the optical conductivity. Additionally, we investigate the effect of a small dipolar symmetry-breaking perturbation on the lattice and the model's crossover behavior at finite temperatures. Finally, in Sec. \ref{Sec:Anisotropy}, we comment on anisotropic cases where the dipole moments are conserved only along selected directions. In the dipole condensed phase of these situations, the low-energy field theory can exhibit fermion-goldstone couplings with directional dependence, which leads to intriguing transport behaviors. We conclude with some remarks on the role of spin and possible physical realizations of these models in Sec. \ref{Sec:conclusion}.

\section{Toy model and Hartree-Fock mean field theory}
\label{Sec:MF}

To write down dipolar symmetric Hamiltonians, let us first define dipole operators $\rho_{\bf{i}}^{(\bf{R})} = \sum_s c^\dag_{\bf{i}+\bf{R},\textit{s}}c_{\bf{i},\textit{s}}$, where $\bf{R}$ is a lattice vector, $s$ can be spin or other internal indices. In the case $s\in \{\uparrow,\downarrow\}$, one can also write down a dipole-invariant bilinears form a spin vector $\vec{d}_{\bf{i}}^{(\bf{R})} =c^\dag_{\bf{i}+\bf{R},\textit{s}}\vec{\sigma}_{ss'}c_{\bf{i},\textit{s}'}$. These operators carry an integer charge $\mathbf{R}\cdot\vec{a}$ under $\hat{Q}_a$. Assuming spinful fermions, we can write down a toy model respecting charge, dipole, and spin symmetry as well as translation symmetry from these dipole operators, which reads
\begin{equation} \label{eq:MicroHamiltonian}
\begin{split}
H &= \sum_{\mathbf{i,j},\mathbf{R}}\mathcal{A}_{\mathbf{ij}}^{\mathbf{R}}(\rho_{\mathbf{i}}^{(\mathbf{R})})^\dag\rho_{\mathbf{j}}^{(\mathbf{R})} + \sum_{\mathbf{i}}\frac{U}{2}n_{\mathbf{i}}^2 +\ldots, 
\end{split}
\end{equation}
where $\mathcal{A}_{\mathbf{ij}}^{\mathbf{R}}$ is a hopping matrix describing the correlated hopping of dipole moments. These are the only processes through which fermions can propagate throughout the lattice. The sum over $\mathbf{R}$ can run over all lattice vectors. For locality, $\mathcal{A}_{\mathbf{ij}}^{\mathbf{R}}$ in general should decay with $|\mathbf{i-j}|$ and $\mathbf{R}$; a hopping matrix for the $\vec{d}$-operators can actually be absorbed into a redefinition of $\mathcal{A}_{\mathbf{ij}}^{\mathbf{R}}$. One can also include other terms that are consistent with the symmetries such as further neighbor density-density interaction. One can view this toy model as a generalized version of the fermi-Hubbard model in a system with dipolar symmetry. For simplicity, we consider the following hopping matrix element which only involve nearest neighbors\cite{LakeDBHM}, 
\begin{equation}\label{eq:LakeHoppingMain}
    \mathcal{A}^{\mathbf{a}}_{\mathbf{ij}} = \sum_{\mathbf{b}}\biggr(-t\delta_{\mathbf{ab}}(\delta_{\mathbf{i,j+a}}+\delta_{\mathbf{i,j-a}}) - t'(1-\delta_{\mathbf{ab}})(\delta_{\mathbf{i,j+b}}+\delta_{\mathbf{i,j-b}})\biggr)
\end{equation}
where $\mathbf{a,b}$ labels the primitive lattice vectors. 

\begin{figure}
    \centering
    \includegraphics[width=0.95\linewidth]{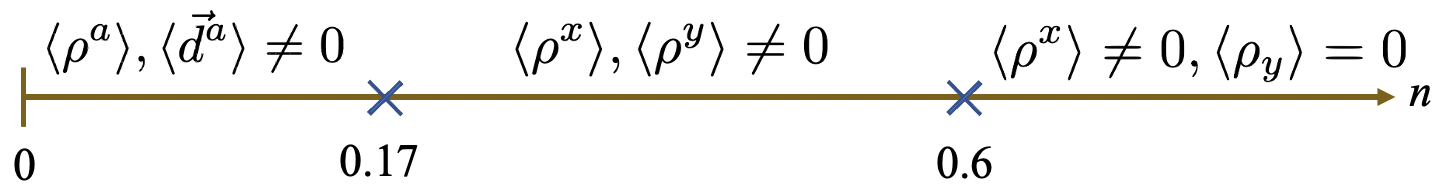}
    \caption{Mean-field phase diagram for the model in Eq. \ref{eq:MicroHamiltonian} with $\tilde{t}<0$, depicting the various phases as a function of the number of fermions per site. The symmetry of the diagram under $n\mapsto 2-n$ is understood.}
    \label{fig:filling_tuned_phase_diagram}
\end{figure}

To illustrate how dipole symmetries may be spontaneously broken, we analyze the Hamiltonian in Eq. \ref{eq:MicroHamiltonian} and \ref{eq:LakeHoppingMain} in 2D without the Hubbard terms, namely the strong dipole hopping limit, at generic fillings. In this limit, it is reasonable to adopt the mean-field ansatz of a dipole symmetry-breaking state where the operators $\rho_i^{\mathbf{R}}$ and/or $\vec{d}_i^{\mathbf{R}}$ acquire a nonzero expectation. We apply a standard Hartree-Fock procedure (see App. \ref{Sec:AppMFTDetails}). 
Within this simple Hartree-Fock treatment, the dipole symmetry is always spontaneously broken in our model, while the specifics of the symmetry-breaking pattern depend on the filling. At low fillings, the mean-field ground state breaks both components of the dipole symmetry and the spin symmetry to form a single ferromagnetic 2D Fermi surface, with all the electrons occupying one spin state. At intermediate fillings, the ferromagnetic order disappears and the system consists of two energetically degenerate 2D Fermi surfaces for each spin state, and at fillings closer to $n=1$ fermion per site, the system prefers to break only one component of the dipole symmetry to generate dispersion in only one direction, forming a quasi-1D state. Finally, we note that depending on the sign of $\tilde{t}=t+(d-1)t'$, the two Fermi surfaces in the non-ferromagnetic phases may be separated in momentum; for $\tilde{t}<0$, they will be completely degenerate, whereas for $\tilde{t}>0$, they are separated by a $(\pi,\pi)$ momentum. These conclusions are summarized in the phase diagram of Fig. \ref{fig:filling_tuned_phase_diagram}, though we note that these conclusions are not exact and may be altered by adding more complicated terms to the simple model Hamiltonian. The purpose of these mean-field results is to illustrate possible diverse dipolar symmetry-breaking phases in these toy models. 

The mean-field results here of course have shortcomings that are common to all mean-field treatments. In fact, fluctuations on top of the mean-field state are very important in these dipolar systems. To study the fluctuations, in the following sections, we will adopt the effective field theory perspective and discuss the universal properties of systems which break one or all components of the dipole symmetry without reference to any particular microscopic Hamiltonian. For most of the paper, we will restrict ourselves to the cases with spinless fermions. The addition of spin does not alter most of the story and will be deferred to the concluding remarks.

\section{Dipole condensation in 1d fermionic systems}
\label{Sec:1D}

Before examining the 2D cases, let us first explore dipolar symmetry breaking in one spatial dimension. Mermin-Wagner theorem prohibits long-range order at zero temperature if we consider dipolar symmetry as a regular $U(1)$ symmetry. However, we demonstrate that coupling to fermions can result in a dipolar symmetry-breaking phase at zero temperature in 1D. This result is consistent with the generalized Mermin-Wagner theorem for multipolar symmetries discussed in Ref. \onlinecite{MultipolarMerminWagner,kapustin_dipole}. Physically, the mobile fermions provide effective long-range interaction between the dipoles which can stabilize the order. The low-energy theory of the 1D model resembles that of the celebrated Schwinger model\cite{Schwinger_Model_OG} and various properties can be obtained exactly through this analogy. 

A useful consistency check for determining whether a symmetry can be spontaneously broken involves assuming a mean-field symmetry-breaking state and analyzing whether the fluctuations of the order parameter have the potential to disrupt the order. We will apply the same logic here. Consider a dipolar symmetric spinless fermion system in 1D and suppose the dipole symmetry has been spontaneously broken. A generic mean-field hamiltonian looks like 
\begin{equation}\label{eq:1dPeierls}
    H_{MF} = \sum_{i}-t Re^{i\varphi} c_{i+1}^\dagger c_i + h.c.
\end{equation}
where $Re^{i\varphi}=\langle\rho^\dagger\rangle$ is the mean-field dipole condensate. The Goldstone mode of the dipole condensate, namely $\varphi$, couples to the fermions exactly like an electromagnetic vector potential. At generic filling, we may pass to the low energy theory of Dirac fermion coupled to the Goldstone mode in a standard way,
\begin{equation}\label{eq:DiracAction}
\mathcal{L}=\bar{\psi}\gamma^\mu(i\partial_\mu-v_F\varphi_\mu) \psi + \mathcal{L}_{NGB},
\end{equation}
where $\psi=(\psi_{R},\psi_L)$, $\{\gamma^0,\gamma^1, \gamma^5\}=\{\sigma_y,i\sigma_x,\sigma_z\}$, $\partial_\mu=(\partial_t,v_F\partial_x)$, and $v_F\sim tRa$ (here $a$ is the lattice spacing). $\mathcal{L}_{NGB}$ is the Lagrangian for the Nambu-Goldstone boson. We also defined $\varphi_\mu = (0, -\varphi)$ in analogy to a gauge field.

Immediately, some low-energy properties follow: due to the axial anomaly of 1D Dirac fermions, the Goldstone propagator is non-perturbatively fixed by the same arguments as in the Schwinger model, which we review below in the context of our theory\cite{GiftsFromAnomalies}. Let us denote $j^\mu=(j^0, j^1)$ (the electromagnetic current) and $j_5^\mu=(j^1,j^0)$ (the axial current) as the Noether currents for $U(1)_V:\,\psi\mapsto e^{i\alpha}\psi$ and $U(1)_A:\,\psi\mapsto e^{i\alpha\gamma^5}\psi$ respectively, normalized such that both would obey $\partial_\mu j_{(5)}^\mu=0$ in the absence of the $\varphi$ coupling. The anomaly equation for the axial current reads\cite{Peskin}

\begin{equation}\label{eq:AnomalyEquation}
\partial_\mu j_5^\mu = -\frac{1}{2\pi}\epsilon_{\mu\nu}F^{\mu\nu}=-\frac{1}{\pi}\partial_t \varphi
\end{equation}
Since $j_5^\mu=-\epsilon^{\mu\nu}j_\nu$ and $\partial_\mu j^\mu = 0$, it follows from Eq. \ref{eq:AnomalyEquation} that 

$$\partial_\mu\partial^\mu j^1=-\frac{1}{\pi}\partial_t^2 \varphi\implies j^1(\omega,k)=-\frac{1}{\pi}\frac{\omega^2}{\omega^2-v_F^2k^2}\varphi.$$ 
If we write the general action for the $\varphi$-field as
\begin{equation}\label{eq:General1DNGBAction}
\mathcal{L}_{NGB}[h]=-\frac{1}{2g^2}\varphi(\partial_t^2-v_B^2\partial_x^2)\varphi - h\varphi
\end{equation}
having added a frequency/wavevector-dependent source term, the equation of motion for $\varphi$ is given by $(D(\omega,k))^{-1}\varphi = h(\omega,k)$, where $D(\omega,k)$ is the exact propagator. Using the equation of motion 

\begin{equation}\label{eq:goldstone_fermion_EOM_1+1d}
\biggr(\frac{1}{g^2}(\partial_t^2-v_B^2\partial_x^2)-v_Fj_1\biggr)\varphi = -h
\end{equation}

and making the substitution for $j_1(\omega,k)$ found earlier, we finally have

\begin{equation}\label{eq:BosonPropNonPert}
D(\omega,k)  = \frac{g^2}{\omega^2-v_B^2k^2 -\frac{v_Fg^2}{\pi}\frac{\omega^2}{\omega^2-v_F^2k^2}}.
\end{equation}

Above, we have proceeded by solving classical equations of motion while including the axial anomaly; we offer a recapitulation of this argument in App. \ref{Sec:AppSchwingerDyson} using the formalism of Schwinger-Dyson equations to emphasize the fact that these manipulations actually take place within the full quantum theory. 

The result in \ref{eq:BosonPropNonPert} is 1-loop exact due to the axial anomaly just as in the Schwinger model. However, there are important differences, for instance, the above Goldstone propagator is not Lorentz invariant and is different from the structure of the gauge propagator in the Schwinger model. Nevertheless, we shall see that many features of the Schwinger model carry over to this system, in particular, the presence of a massive pole in the boson spectrum and the `confinement' of fermions. 

We can represent the Goldstone propagator in the following form which manifests the pole structure:
\begin{equation}\label{eq:PropagatorSpectralForm}
\begin{split}
    D(\omega,k)  &= g^2\biggr(\frac{\mathcal{Z}_p(k)}{\omega^2-\omega_p^2(k)}+\frac{\mathcal{Z}_g(k)}{\omega^2-\omega_g^2(k)}\biggr).
\end{split}
\end{equation}
The pole positions for small momentum $k^2\ll g^2/\pi v_F$ are given by:
\begin{equation}\label{eq:Poles1DGB}
\begin{split}
    \omega^2_{p}(k) &= m^2+(v_F^2+v_B^2)k^2 + \ldots\\
    \omega^2_{g}(k) &= \frac{2v_F^2v_B^2}{m^2}k^4 + \ldots,\\
\end{split}
\end{equation}
where $m^2=v_Fg^2/\pi$. And the spectrum weight is given by
\begin{equation}
\begin{split}
\mathcal{Z}_p(k) &= \frac12\biggr(1 + \frac{m^2 -(v_F^2-v_B^2)k^2}{\sqrt{((v_F^2+v_B^2)k^2+m^2)^2-4v_B^2v_F^2k^4}}\biggr)\\
    &\cong 1-\frac{v_F^2k^2}{m^2}+...\\
    \mathcal{Z}_g(k) &= \frac12\biggr(1 - \frac{m^2 -(v_F^2-v_B^2)k^2}{\sqrt{((v_F^2+v_B^2)k^2+m^2)^2-4v_B^2v_F^2k^4}}\biggr)\\
    &\cong \frac{v_F^2k^2}{m^2}+...
    \end{split}
\end{equation}
Thus, due to the coupling with the fermions, the spectral weight of the free Goldstone boson is redistributed to a massive, plasmon-like pole and a massless pole with a quadratic dispersion (indicating a dynamical exponent $z=2$). However, the spectral weight of the massless pole $\mathcal{Z}_g(k)$ vanishes as $k\to 0$. Accordingly, the integral $\int d\omega dk D(i\omega,k)$ is \textit{not} IR-divergent. This suggests that the long-range order for the dipole condensate can be stable in a (1+1)-d system at zero temperature, as opposed to the quasi-long-range order one would na\"ively expect. 

One can also understand the low-energy physics of this 1D system from a bosonization perspective\cite{kogut_sinclair,kogut_susskind,casher_kogut_susskind,fradkin_2013}, where we may naturally include fermion interactions $\mathcal{H}_{int} = \frac{\pi}{2}(V^\phi j_0^2 + V^\theta j_1^2)$\footnote{Surprisingly, the spatial current $j_1$ is a dipole-invariant operator; this follows from the perfectly linear dispersion of the Dirac fermion. Quadratic and higher corrections are proportional to 2nd and higher derivatives of the dispersion near $k_F$, and are thus irrelevant perturbations to the Luttinger liquid action. One may ask about the interaction terms involving $\varphi$ generated by such corrections to the dispersion, but upon noticing that the coupling $g$ becomes a mass scale, $\varphi$ also acquires dimensions of mass, and thus all such interaction terms are rendered irrelevant}. With the  convention $\psi_R\sim e^{i\sqrt{\pi}(\vartheta+\phi)}, \psi_L\sim e^{i\sqrt{\pi}(\vartheta-\phi)}$, the low energy theory in terms of bosonic variables can be written as the following:
\begin{equation}\label{eq:1DLLactionPhi}
    \mathcal{S}_\phi = \frac{\kappa}{2}\int dt dx \biggr(\frac{1}{v}(\partial_t\phi-\frac{v}{\kappa}\varphi_x/\sqrt{\pi})^2-v(\partial_x\phi)^2 - \frac{v}{\kappa^2\pi}\varphi_x^2 \biggr) 
\end{equation}
or
\begin{equation}\label{eq:1DLLactionTheta}
    \mathcal{S}_\vartheta = \frac{1}{2 \kappa}\int dt dx \biggr(\frac{1}{v}(\partial_t\vartheta)^2-v(\partial_x\vartheta-\varphi_x/\sqrt{\pi})^2\biggr) 
\end{equation}
with the action $\mathcal{S}_{NGB}$ implicit. Here, $v=\sqrt{(v_F+V^\phi)(v_F+V^\theta)}$ and $\kappa=\sqrt{(v_F+V^\phi)/(v_F+V^\theta)}$ are the usual Luttinger parameters. The action in Eq. \ref{eq:1DLLactionTheta} was studied in Ref. \onlinecite{LakeDBHM, Lake_Tilted_Chain, zechmann2022fractonic} in the context of dipole-conserving bosons, where $\vartheta$ plays the role of the superfluid phase mode. Here, the microscopic origin of $\vartheta$ is slightly different as it represents the phase mode of the right/left moving fermions, but it still transforms by a linear polynomial shift under the combined charge and dipole symmetries. One can shift $\varphi_x\to \varphi_x+\sqrt{\pi}\partial_x\vartheta$ in Eq. \ref{eq:1DLLactionTheta}, gapping out the $\varphi$-field and turning $\vartheta$ into a quadratically dispersing field; in fact, this dispersion relation for $\vartheta$ manifests in the pole $\omega_g^2(k)$ found in Eq. \ref{eq:Poles1DGB}. 

\begin{figure}
    \centering
    \includegraphics[width=0.8\linewidth]{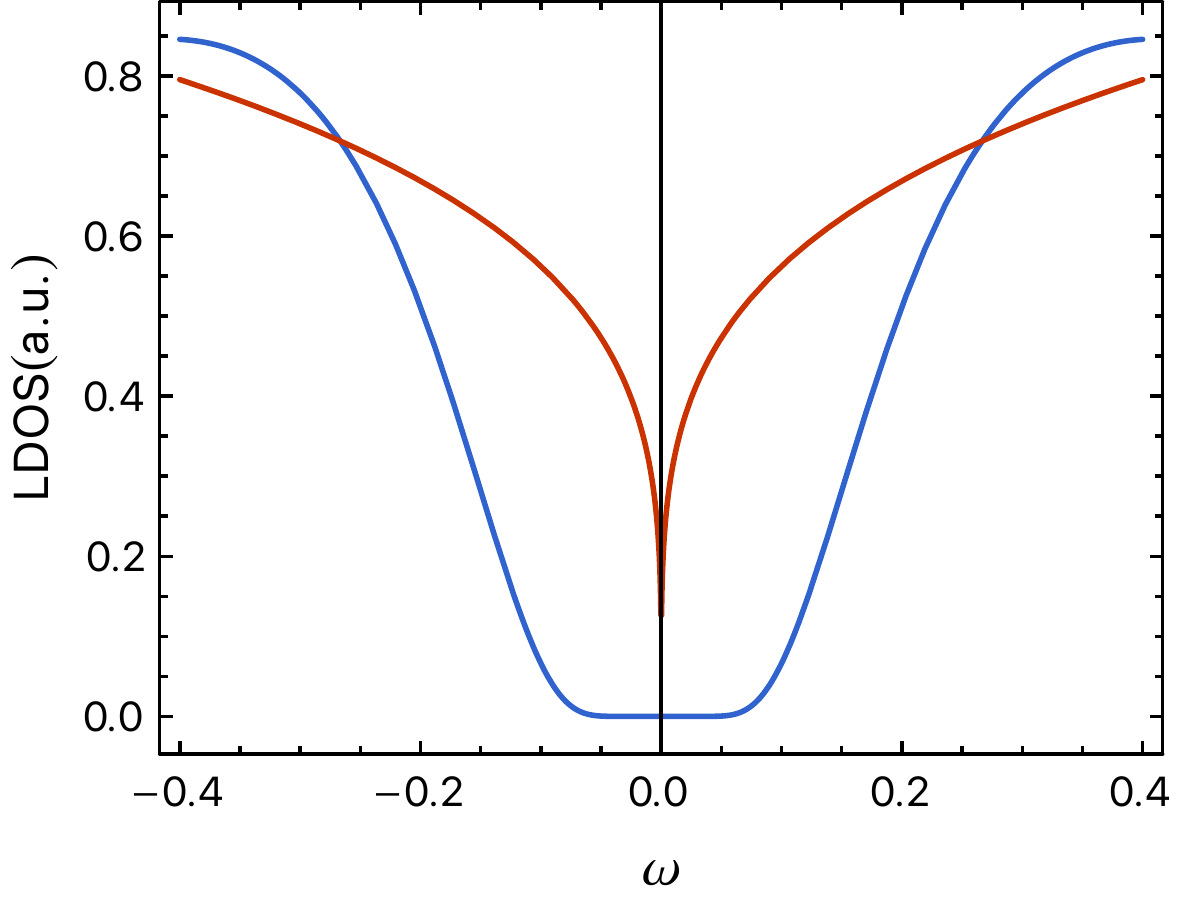}
    \caption{Low-frequency local density of states for a spinless Luttinger Liquid with $\kappa=2$ (red) and the dipole-symmetric model in Eq. \ref{eq:1dPeierls} (blue).}
    \label{fig:LDOS}
\end{figure}

The bosonized action can be used to easily compute correlation functions of fermions with interaction. The details are worked out in App. \ref{Sec:AppBosonization}. It is shown that $\avg{\psi_R(x,0)\psi_R(0,0)^\dagger}\sim e^{-\abs{x}/\xi}$ for $x\gg \xi$, where $\xi^{-1}=m\kappa/(4\pi v_B)$. Hence, single fermions are `confined' in analogy to the Schwinger model. Furthermore, at long times, $\avg{\psi_R(0,t)\psi^\dag_R(0)}\sim e^{-(\pi/\sqrt{2})\sqrt{vt/\xi}}$, and the one-particle local density of states may be computed for low frequencies and is seen to scale as $\abs{\omega}^{-3/2}\exp(-\frac{\pi^2}{16}\frac{v}{\xi\abs{\omega}})$, i.e. the density of states near $\omega=0$ has an essential singularity, shown in Fig. \ref{fig:LDOS}. It is useful to contrast these results to the case of an ordinary spinless Luttinger liquid, where $\avg{\psi_R(x,0)\psi^\dag_R(0)}\sim \abs{x}^{-\frac12(\kappa+\kappa^{-1})}$ and the local density of states scales as a power law $\abs{\omega}^{\frac12(\kappa+\kappa^{-1})-1}$ \cite{fradkin_2013}. In terms of the microscopic mean-field parameters, one can show that $v_B^2 \sim tR^2a$ and $g = a^{-1}$. Hence, the mass scale $m^2 \sim tR/a$ and $\xi \sim a\sqrt{R}$, i.e. the confinement length scale is on the order of the lattice spacing.

We briefly comment on the issue of stability. It has been noted in \cite{Lake_Tilted_Chain, zechmann2022fractonic} that the phase described by Eqs. \ref{eq:1DLLactionPhi}-\ref{eq:1DLLactionTheta} exhibits LRO in the correlator $\avg{e^{i2\sqrt{\pi}\phi(x)}e^{-i2\sqrt{\pi}\phi(0)}}$, hence a term $\cos(2\sqrt{\pi}\phi)$ that would pin $\phi = 0$ is highly relevant. Since $\cos(2\sqrt{\pi}\phi)$ is mapped to $\bar{\psi}\psi$ in the fermion language in Eq. \ref{eq:DiracAction}, this is the indication that the Dirac theory coupled to the dipolar Goldstone mode is unstable to chiral symmetry breaking, and microscopically the chiral mass term should be mapped to certain $2k_F$ charge-density wave (CDW) order parameter. Addtionally, we can see the chiral symmetry breaking more explicitly in the low energy theory upon integrating out the dipolar Goldstone mode $\varphi$ in Eq. \ref{eq:1DLLactionPhi}. Similar to the Schwinger model, integrating out the Goldstone modes generates a mass term for $\phi$ in the long wavelength limit, hence the CDW order parameter takes a nonzero expectation value. However, numerical studies may ultimately be needed to verify the predictions of this IR field theory, in particular whether the true dipolar LRO can be observed in 1-D fermionic systems.

To summarize, we have shown that imposing dipole symmetry on a Luttinger liquid in 1+1d drastically changes its low-energy properties, many of which may be understood within the dual picture of the 'Bose-Einstein Insulator' phase of dipole-conserving bosons in 1D. The 'confinement' of single fermions follows from the dipole symmetry in a generic manner; such a connection between these two phenomena has been appreciated before in other 1D models \cite{pretko_1d_confinement_fractons}. We have restricted this discussion to zero temperature; at finite temperature, one may check that $\sum_{i\omega_n,k_x}D(i\omega_n,k_x)$ is IR divergent, so there is no dipolar LRO or QLRO.

\section{Dipole condensation in $2d$ and non-Fermi liquid states}
\label{Sec:2D}
\subsection{Low energy theory}

Our starting point is a mean-field state which preserves the charge $U(1)$ symmetry but spontaneously breaks the dipolar symmetry along all directions, leading to dispersive fermion bands which can be described in general by the following Bloch Hamiltonian:
\begin{equation} \label{eq:GeneralMeanFieldHami}
H_0 = \sum_{\vec{k},s,s'}c^\dag_{k,s}h_{s,s'}(\vec{k})c_{k,s'}
\end{equation}
The indices $s$ can be the spin or other indices; here we treat it as a general internal degree of freedom, and the index $s$ may range from $1$ to $N$. At generic fillings, the mean-field state hosts fermi surfaces. In general, we will assume that the shape of the fermi surface is sufficiently regular and that the Bloch eigenvectors have no momentum dependence.

The fluctuations beyond the mean-field come from coupling to the Goldstone bosons of the broken symmetries. Assuming the mean-field state spontaneously breaks dipolar symmetry along all directions, this gives us $d$ Goldstone bosons. There could be other Goldstone modes for internal symmetry breaking. For instance, spin symmetry could be broken and we will discuss their consequences in Sec. \ref{Sec:conclusion}. The influence of Goldstone modes on the low-energy fermions may vary depending on the structure of the broken symmetry. 

It was shown in Ref. \onlinecite{CriterionVishwanath} that the Goldstone modes whose associated symmetries do not commute with translation have nonvanishing couplings to the fermi surface at zero momentum. Here, the dipolar symmetries in fact satisfy this criterion, hence their Goldstone modes can significantly alter the low-energy physics. We can compute the dipolar Goldstone-fermion interactions given only the form of the mean-field Hamiltonian and the microscopic representation of the dipole charge operators (see App. \ref{Sec:AppGoldstoneCouplings} for more details). The resulting low-energy theory including low-momentum dipolar Goldstone fluctuations takes the following form:

\begin{equation} \label{eq:MicroscopicAction}
\begin{split}
\mathcal{S} &= \mathcal{S}_{f} + \mathcal{S}_{GB}+\mathcal{S}_{int}\\
\mathcal{S}_{f} &=\sum_s\int d\tau\frac{d^2k}{(2\pi)^2} \bar{\psi}_{\vec{k},s}(\tau)\biggr(\partial_\tau+\xi_{\vec{k}}\biggr) \psi_{\vec{k},s}(\tau) + \ldots\\
\mathcal{S}_{GB} &= \int d\tau d^2x\frac{1}{2g^2}\biggr(\sum_a(\partial_\tau\varphi^a)^2+v_1^2\sum_{ab}(\partial_a\varphi^b)^2  \\ &+v_2^2(\sum_a\partial_a\varphi^a)^2+v_3^2\sum_a(\partial_a\varphi^a)^2\biggr)\\
\mathcal{S}_{int} &=\sum_{k,q,s}\sum_a\varphi^a_q\bar{\psi}_{k+q/2,s}\psi_{k-q/2,s}\partial_a\xi_{\vec{k}}\\
&+\frac{1}{2}\sum_{k,q,q',s}\sum_a\varphi^a_{q}\varphi^b_{q'}\bar{\psi}_{k+(q+q')/2,s}\psi_{k-(q+q')/2,s}\partial_a\partial_b\xi_{\vec{k}}
\end{split}
\end{equation}
where $\partial_a = \hat{a}\cdot\grad_k$, $\xi_{\vec{k}}=\epsilon_{\vec{k}}-\mu$ is the dispersion generated by dipole condensation, $s \in\{1,\ldots,N\}$, and $g>0$ is a coupling strength. The form of the NGB action is fixed by the symmetries -- the only possible linear in $\partial_\tau$ term, $\epsilon_{ab}\varphi^a\partial_\tau\varphi^b$, is excluded because the dipolar symmetries along different directions commute with each other, $[\hat{Q}_a,\hat{Q}_b] =0$ \cite{Watanabe_Goldstone_Theorem}. For the remainder of this work, we assume the term proportional to $v_3^2$ (arising due to cubic anisotropy) vanishes; we discuss the effects of such a term in App. \ref{Sec:AppAnisotropy}, none of which are relevant to describing the IR behavior of this model. We include the coupling between fermions and Goldstone modes up to the second order in the Goldstone field. Higher order coupling terms are given by higher derivatives of the Bloch Hamiltonian, and we may also have additional operators in the Hamiltonian that are invariant under the dipole symmetry (e.g. a residual on-site interaction); in the following treatments, we neglect such terms, as they are irrelevant perturbations to the low-energy theory. Notably, the Goldstone modes couple to the fermions in a manner that is very similar to the spatial part of a $U(1)$ gauge potential. This is easy to understand: as we argued in the 1D case around Eq. \ref{eq:1dPeierls}, the coupling of the Goldstone in the microscopic lattice model resembles a Peierls substitution for a tight-binding Hamiltonian. We also note that the Hamiltonian for the Goldstone fields does not enjoy any gauge invariance; there is no analog of the temporal component $A^0$ in the Goldstone fields, and the Goldstone action is generally not invariant under the transformation $\varphi^a\mapsto \varphi^a-\partial^a\chi$. 

A standard calculation of the Goldstone self-energy in 2D at RPA level gives the following result, 
\begin{equation}
\label{eq:OneLoopSelfEnergy}
   \Pi_{ab}(i\omega,\vec{p}) \approx -\gamma \frac{\abs{\omega}}{\abs{\vec{p}}}(\delta_{ab}-p_ap_b/\abs{\vec{p}}^2)
\end{equation}
where $\gamma=\frac{Nv_F}{4\pi\kappa}$, and $v_F$ and $\kappa$ are the fermi velocity and the curvature of fermi surface respectively. Correspondingly the Goldstone propagator is given by 
\begin{equation}\label{eq:NGBPropagatorMainText}
\begin{split}
D_{ab}(i\omega,\vec{p})&= \frac{g^2(\delta_{ab}-p_ap_b/\abs{\vec{p}}^2)}{\omega^2+v_1^2\abs{\vec{p}}^2+\gamma g^2\frac{\abs{\omega}}{\abs{\vec{p}}}} + \frac{g^2{p_ap_b}/{\abs{\vec{p}}^2}}{\omega^2+(v_1^2+v_2^2)\abs{\vec{p}}^2}\\
&=D^T_{ab} + D^L_{ab}.
\end{split}
\end{equation}
We can define the longitudinal $\varphi_L(p)$ and transverse $\varphi_T(p)$ component of the Goldstone mode as the component of $\vec{\varphi}(p)$ parallel and perpendicular to $\vec{p}$ respectively. The propagator in Eq. \ref{eq:NGBPropagatorMainText} indicates that the longitudinal propagator receives no correction at one loop while the transverse propagator receives Landau-damping.

We then consider the effect of Goldstone coupling on the fermion self-energy. Due to energetic constraints, the Goldstone field most strongly couples to fermions wherever its momentum is parallel to the fermi surface. Since only the transverse mode gets damped, we will see that as a result, only $\varphi_T(\vec{p})$ couples strongly to the Fermi surface within the patch decomposition. We now pass the previous action to the patch decomposition, which has been utilized extensively in the past for 2D models of fermi surfaces coupled to critical bosons\cite{sachdevQPT,sslee2009,metlitski_sachdev_ising,Max_Cooper_Pairing_NFL}. In App. \ref{Sec:AppPatchTheory}, we argue that after discarding all irrelevant terms in the patch action, the low-energy theory of two antipodal patches on the Fermi surface with the Goldstone boson is
\begin{equation}\label{eq:PatchLagrangianMainText}
    \begin{split}
        \mathcal{S} &= \mathcal{S}_{GB} + \sum_{\theta\in\pm}\mathcal{S}_{\theta} \\
\mathcal{S}_{GB} &= \int d\tau d^2x\biggr(\sum_{i\in \{T,L\}}\frac{1}{2g^2}v_i^2(\partial_y\varphi_i)^2 \biggr) \\
\mathcal{S}_{\theta} &=\sum_s\int d\tau d^2x\, \bar{\psi}_{\theta,s}\biggr(\eta \partial_\tau + s_\theta v_F(i\partial_x +\varphi_T) - \frac12\kappa \partial_y^2\biggr)\psi_{\theta,s}
    \end{split}
\end{equation}
where $v_T^2=v_1^2$ and $v_L^2 = v_1^2+v_2^2$ are the bare propagation speeds of the transverse and longitudinal Goldstones, and $x,y$ respectively refer to the parallel and perpendicular directions to the patches. Aside from the propagating $\varphi_L$ field, which decouples from the patches in the IR, this two-patch action is the same as that of the IR theory of a Fermi surface coupled to a transverse $U(1)$ gauge field. We note again that there is no gauge symmetry in our setup -- the longitudinal degree of freedom $\varphi_L$, while decoupled from the fermi surface, is still a propagating mode. It is interesting that while the dipolar Goldstone always couples like a gauge field to the matter fields, it is the dynamics of the fermi surface that eventually causes the low-energy theory to resemble the structure of a gauge theory by decoupling the longitudinal mode in the IR\footnote{Strictly speaking, tuning $v_2^2=-v_1^2$ in the free Goldstone action would also lead to a gauge-theoretic structure by making the action invariant under $\varphi^a\mapsto \varphi^a-\partial^a\chi(x)$. However, this is an extremely fine-tuned condition.}. 

The above argument in fact holds for any dimension $d\geq2$. For example, in Appendix \ref{Sec:AppBosonSE}, we show that in 3D, the Goldstone self-energy is also proportional to $\delta^{ab}-p_ap_b/|\vec{p}|^2$, implying that the two transverse Goldstone modes are Landau damped while the remaining longitudinal mode propagates freely. In general, when all $d$ components of the dipole symmetry are spontaneously broken, the IR theory is isomorphic to the IR patch theory of a $d$-dimensional Fermi surface coupled to a $U(1)$ gauge field with an additional decoupled longitudinal mode. Ultimately, the observation that the coupling to the longitudinal mode is always less relevant than that of the transverse modes comes from the scaling relation $[p_x]=2[p_y]$ between momenta perpendicular/parallel to the Fermi surface patch. 

From the patch theory, we can show (see App. \ref{Sec:AppFermionSE} for details) that the one-loop self-energy of the fermions comes from the interaction with $\varphi_T(\vec{p})$ and is given by
\begin{equation} \label{eq:OneLoopSelfEnergyFermion}
\begin{split}
\Sigma(i\omega,\vec{q}) &= -i\sign(\omega)\biggr(\frac{1}{6\sqrt{3}\pi^2}\frac{g^4v_F^2 \kappa}{Nv_1^4}\biggr)^{1/3}\abs{\omega}^{2/3}\\
&=-i\sign(\omega)E_{NFL}^{1/3}\abs{\omega}^{2/3}
\end{split}
\end{equation}
for $\omega\ll E_{NFL}$, and $\vec{q}$ on the Fermi surface. This suggests that in 2D quasiparticles near the Fermi surface are destroyed. The generality of this result suggests that in 2 dimensions any Fermi surface ground state coming from spontaneously broken dipole symmetry is automatically a non-Fermi liquid (NFL) at low energies. The phenomenon of Landau-damped dipolar Goldstone bosons leading to a destabilization of the Fermi surface\cite{oganesyan_kivelson_fradkin_goldstone_NFL, ruhman_berg_goldstone_NFL, bahri_potter_goldstone_NFL,mandal_boson_mass_NFL_crossover} is a new example within the general criterion of Ref. \onlinecite{CriterionVishwanath}. 
Per the results of Ref. \onlinecite{Max_Cooper_Pairing_NFL}, we expect that this non-Fermi Liquid generally remains stable to superconducting instabilities. In the following section, we will use the scaling dimensions at this NFL fixed point to make some more claims about the system when dipolar symmetry is slightly explicitly broken.

\subsection{Approximate dipole symmetry and crossover}

Before considering explicit symmetry breaking, let us discuss the energy scale in the symmetric case. In 2 spatial dimensions, there are ostensibly two temperature scales present in Eq. \ref{eq:MicroscopicAction}—the BKT transition temperature $T_{BKT}\sim v_1^2$ above which the dipole order is destroyed and the theory in Eq. \ref{eq:MicroscopicAction} is no longer valid, and the temperature $T_{NFL}\sim E_{NFL}$ above which the system's behavior smoothly crosses over to that of a Fermi liquid. However, due to the shared origin of the Fermi velocity and the stiffness parameter $v_1^2$ via dipole condensation, these microscopic scales are parametrically related. Via the mean-field theory, we can estimate $v_1^2\sim t\abs{R}^2$, $v_F\sim t\abs{R}a$, $\kappa\sim t\abs{R}a^2$, and $g=a^{-1}$ (here $t$ is a characteristic hopping amplitude for dipoles, and $a$ is the lattice spacing). Hence, in terms of the dipolar condensate and the microscopic parameters, we have
\begin{equation}\label{eq:BKT_NFL_Scales}
    \begin{split}
    T_{BKT}&\sim t\abs{R}^2\\
    T_{NFL}&\sim \frac{t}{\abs{R}}
    \end{split}
\end{equation}

We note that $\abs{R} = O(1)$ within a Hartree-Fock treatment of our model \ref{eq:MicroHamiltonian}. Therefore we conclude that $T_{NFL}$ and $T_{BKT}$ are set by the same energy scale, up to non-universal dimensionless constants that depend on (e.g.) the filling, microscopic Hamiltonian, and the precise nature of the dipole ordering. Thus, below $T_{BKT}\sim T_{NFL}$ the NFL state of our discussion takes place.

We are interested in the effect of slightly breaking the dipolar symmetry, i.e. by explicitly adding any small perturbation to the microscopic Hamiltonian that does not commute with the dipole moment $Q_a$. Such a situation of approximate dipole conservation is particularly relevant for experimental implementations involving tilted optical lattices, as the dipole symmetry is only typically preserved up to higher-order terms \cite{Lake_Tilted_Chain}. In the low energy theory, this will be reflected by a small mass $r_0\varphi^2$ for the Goldstone modes, namely, the Goldstones $\varphi^a$ will become \textit{pseudo}-Goldstone bosons. The Goldstone mass is apparently a relevant perturbation to the NFL fixed point and lower than a certain crossover temperature the system will become a fermi liquid due to this perturbation.  

\begin{figure}
    \centering
   \includegraphics[width=0.8\linewidth]{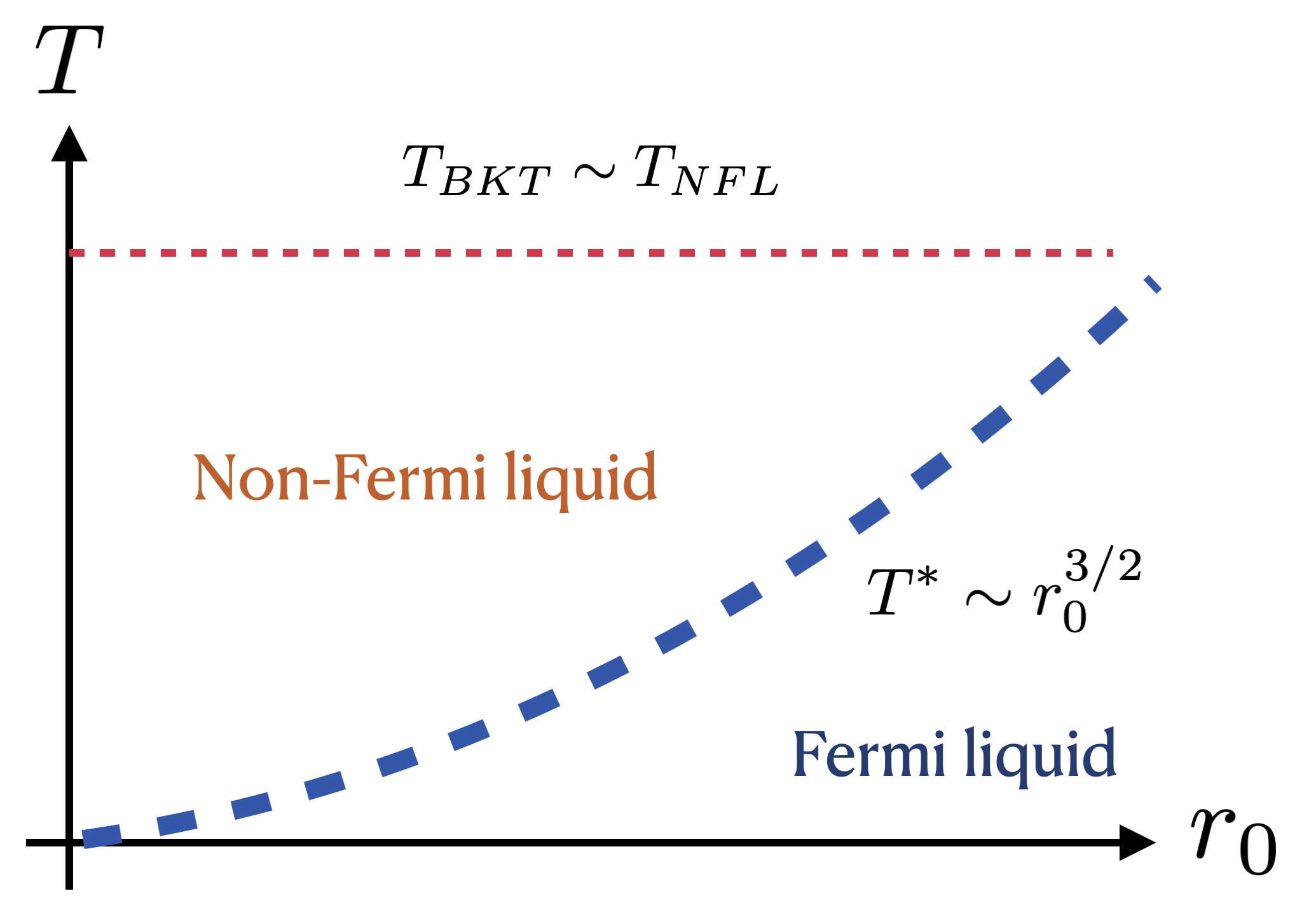}
   \caption{Schematics of finite temperature crossover.}
    \label{fig:Crossover_Sketch}
\end{figure}

To determine the crossover temperature, we adopt the scaling relation from the leading order in a combined $\epsilon$ and large $N$ expansion\cite{Max_Cooper_Pairing_NFL, nayak_wilczek_nFL_fixed_point,nayak_wilczek_nFLRG, mross_mcgreevy_liu_senthil_NFL_eps_N}. The theory without the Goldstone mass is expected to flow to a NFL fixed point where the boson field scales as $\varphi_T'(x',y',t') = b^{2/3}\varphi_T(x,y,t)$ with $\tau' = b^{-1}\tau$, $x'=b^{-2/3}x$, $y'=b^{-1/3}y$. Within the patch decomposition, the Goldstone mass does not get renormalized at one loop\cite{metlitski_sachdev_ising,mross_mcgreevy_liu_senthil_NFL_eps_N}, so the scaling $r\mapsto b^{2/3}r'$ may be directly read off from the above scaling dimensions at the fixed point. The crossover energy/temperature scale is roughly the scale where the $r$ parameter flows to $O(1)$. Therefore, we expect a crossover temperature $T^* \sim r_0^{3/2}$ below which the system behaves like a fermi liquid as demonstrated schematically in Fig. \ref{fig:Crossover_Sketch}. 

The same argument can be made more explicit by calculating the one-loop fermion self-energy in the presence of a finite Goldstone mass,
\begin{equation}\label{eq:MassiveGoldstoneFermiSEScalingMaintext}
 \abs{\Sigma(i\omega)}\approx
  \begin{cases}
                                   c_r(E_{NFL}/\omega^*)^{1/3}\omega & \text{\,\,$\omega\ll \omega^*$} \\
                                   E_{NFL}^{1/3}\omega^{2/3} & \text{\,\,$\omega^*\ll\omega\ll E_{NFL}$} 
  \end{cases}
\end{equation}
where 
\begin{equation}\label{eq:nFLCrossoverTemp}
    \omega^* = (Ng^2v_1v_F/\kappa)^{-1}r_0^{3/2} \sim T^*,
\end{equation}
and $c_r\approx 0.55$ is found numerically. The crossover frequency scale $\omega^*$ is identified with the crossover temperature $T^*$. It has been shown that similar theories of Fermi surfaces with a Yukawa coupling to an order parameter near criticality\cite{lohneysen_HMM_theory,sachdevQPT,optical_conductivity_2dQCP_Sachdev,mandal_boson_mass_NFL_crossover} exhibit the same crossover energy scale proportional to $r_0^{3/2}$. This crossover scale will manifest in the optical conductivity of our model which is considered next. 

\subsection{Optical conductivity}
We compute the optical conductivity of our theory in the standard way: couple the theory to a probe vector potential $\vec{A}$, integrate out the fermions and NGBs, and read off the quadratic piece $\frac12A_a(\omega,\vec{q})K_{ab}(\omega,\vec{q})A_b(-\omega,-\vec{q})$ in the effective action. We note that this will capture both the paramagnetic and diamagnetic contributions to the current. Upon noticing that the vector of NGBs $\vec{\varphi} = (\varphi^x,\varphi^y)$ couples to the fermions in the same way as the vector potential, we may simply couple to $\vec{A}$ by replacing $\vec{\varphi}\to \vec{\varphi}-\vec{A}$ in the coupling terms. 

At the RPA level, it follows that the conductivity for the full theory including the Goldstone-Fermi interactions is given by
\begin{equation}\label{eq:RPAconductivityCancelling}
\begin{split}
    K_{ab}(\vec{q},i\omega)  &= K_{ab}^0(\vec{q},i\omega) + K_{ac}^0(\vec{q},i\omega)D_{cd}(\vec{q},i\omega)K^0_{db}(\vec{q},\vec{i\omega})\\
 &= K_{ab}^0 + K_{ac}^0\frac{1}{(D^0)^{-1}_{cd}-K^0_{cd}}K^0_{db}
\end{split}
\end{equation}
where $D_{cd}(\vec{q},i\omega)$ is the RPA boson propagator in Eq. \ref{eq:NGBPropagatorMainText}, $K_{ab}^0$ is the Drude contribution (i.e. the non-interacting current-current correlator plus the diamagnetic piece), $D^0$ is the free Goldstone propagator, and we recognize $\Pi_{ab}(\vec{q},i{\omega})=K_{ab}^0(\vec{q},i\omega)$. As we take $\vec{q}\to 0$, 

\begin{equation}\label{eq:RPAconductivityCancelling2}
\begin{split}
\mathbf{K}(0,i\omega)  &= \mathbf{K}^0(0,i\omega)\biggr(\mathbbm{1} - \frac{1}{-\frac{1}{g^2}\omega^2(\mathbf{K}^0(0,i\omega))^{-1}+\mathbbm{1}}\biggr)\\
&= O(\omega^2)
\end{split}
\end{equation}
In fact, it has been shown \textit{nonperturbatively} that this expression for the optical conductivity is exact in a patch theory similar to the one discussed in \ref{Sec:2D} due to anomaly considerations\cite{ersatzFL,GiftsFromAnomalies,shi2022loop}. Moreover, this nonperturbative calculation goes through even when the NGBs are given a finite mass. An explicit expression for $\mathbf{K}^0(0,i\omega)$ is given in Eq. \ref{eq:SEDrudeWeight}; per Ref. \onlinecite{GiftsFromAnomalies}, the one-loop relation $\Pi_{ab}(0,i{\omega})=K_{ab}^0(0,i\omega)$ is also exact in the patch theory.

In the limit of $\omega\to 0$, we see that $K_{ab}(\vec{q},\omega)$ is screened to zero, and the coefficient of the $\omega^{-1}$ part of the optical conductivity thus vanishes. This is not particularly surprising. In fact, a stronger statement that the $\vec{q}=0$ conductivity \textit{must} vanish in a dipole-conserving system at all frequencies can be shown easily by considering the commutator between the Hamiltonian and dipolar charge operator\cite{LakeDBHM}. Eq. \ref{eq:RPAconductivityCancelling} is consistent with this general result. It appears to require an IR limit for the conductivity to vanish only because we have implicitly been working within a low-energy theory from the start—the apparent nonzero conductivity at $\vec{q}=0$ comes from the term $(\partial_\tau\varphi^a)$ in the Goldstone action \ref{eq:MicroscopicAction}, which is in principle generated by integrating out high-energy fermions. Such degrees of freedom manifest as irrelevant contributions to the total current operator that are neglected in our treatment. It will be interesting to check in the future if we can get exact zero optical conductivity with more comprehensive considerations. Nevertheless, this analysis captures the correct IR behavior of our system, and we will use Eq. \ref{eq:RPAconductivityCancelling} as a starting point to discuss the crossover behavior.

\begin{figure}[h]
    \includegraphics[width=0.8\linewidth]{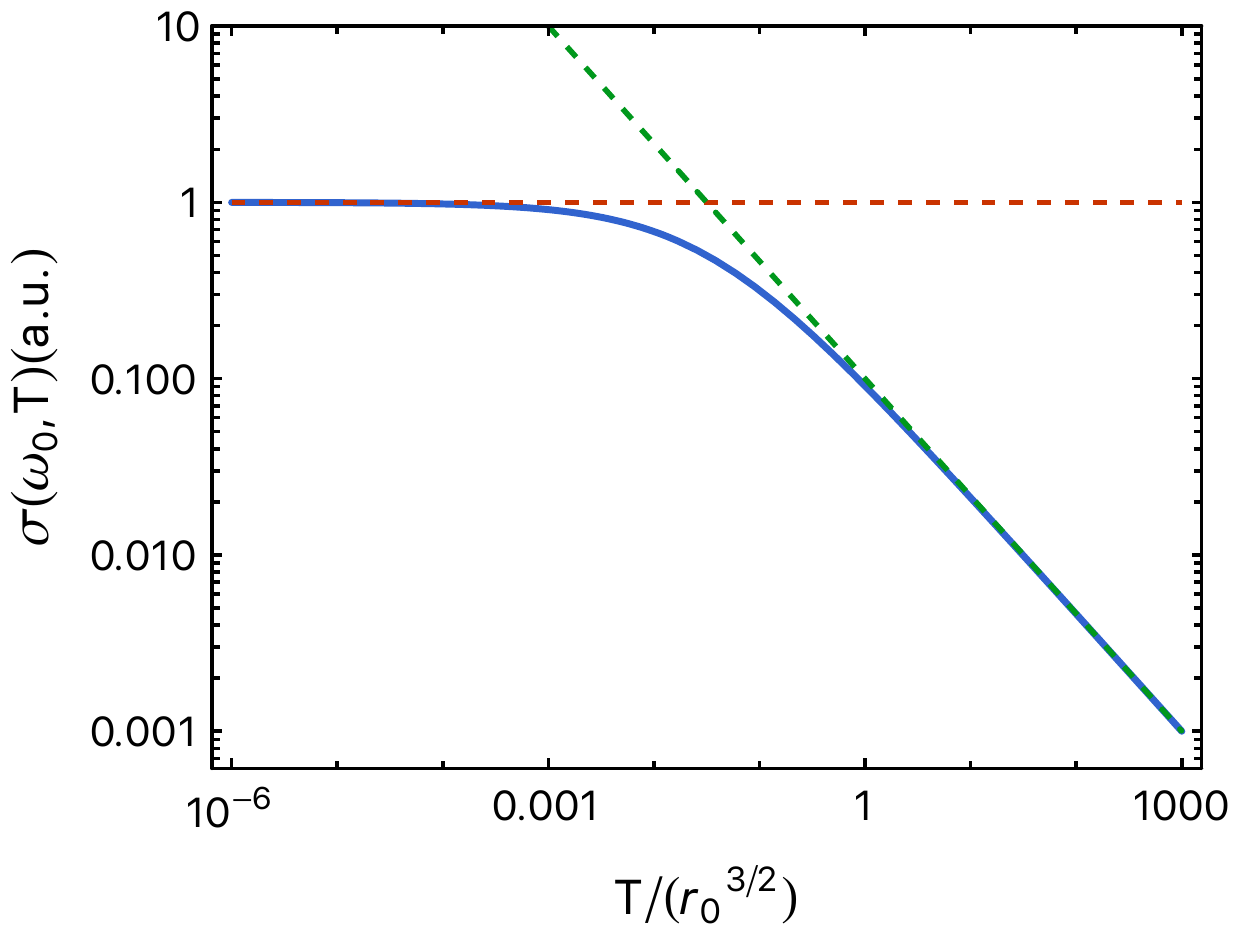}
    \caption{A schematic plot for optical conductivity as a function of temperature at a fixed frequency. Dashed lines indicate $T^0$ and $T^{-2/3}$ dependence at low and high temperatures respectively.}
    \label{fig:conductivity_crossover}
\end{figure}

If we suppose that the dipole symmetry is slightly broken, then the Goldstones gain a finite mass. In the presence of a Goldstone mass-matrix $\frac12 r_{ab}\varphi_a\varphi_b$, recapitulating the earlier computation (in real-time) for the low-frequency conductivity implies
\begin{equation}\label{eq:ApproxDrudeWeight}
    \begin{split}
        \mathbf{K}(0,\omega)  &= \mathbf{K}^0(0,\omega) - \mathbf{K}^0(0,\omega)\frac{1}{1+\frac{1}{g^2}\mathbf{r}(\mathbf{K}^0)^{-1}(0,\omega)}\\
 \implies \sigma(\omega) &= -\frac{i}{\omega}\frac{1}{g^2}\biggr( \frac{\mathbf{K}^0(0,\omega)\mathbf{r}(\mathbf{K}^0)^{-1}(0,\omega)}{1+\frac{1}{g^2}\mathbf{r}(\mathbf{K}^0)^{-1}(0,\omega)}\biggr) .
    \end{split}
\end{equation}
Thus, the explicit breaking of dipole symmetry is seen to restore a finite optical conductivity, which takes the form of a Drude peak with weight set by the Goldstone mass matrix (to leading order in $\mathbf{r}$). Taking $\mathbf{r} = r\mathbbm{1}$ for simplicity, from the scaling argument in the previous section, we have $r(T)=r_0T^{-2/3}$ provided $T$ is large enough to keep $r(T)$ close to the nFL fixed point. Thus, the Drude peak in Eq. \ref{eq:ApproxDrudeWeight} will scale with temperature as $T^{-2/3}$ at high temperatures, whereas at low temperatures, the Goldstone mass runs to large values and a na\"ive interpretation of Eq. \ref{eq:ApproxDrudeWeight} suggests that the fixed frequency conductivity will be constant in $T$. Of course, we cannot trust the linearized RG flow at a fixed point once a relevant parameter has flowed far away from the fixed point. Still, the prediction of a constant-in-$T$ Drude weight for $T\ll T^*$ makes sense physically, as the Goldstone fluctuations will surely be gapped out leaving behind only a Fermi liquid as the relevant piece of the theory. This scaling behavior is schematically sketched in Fig. \ref{fig:conductivity_crossover}. We emphasize that we have only discussed the "pole" part of the optical conductivity, neglecting other potential terms proportional to $\omega^{-2/3}$ in quantum critical systems, as suggested in e.g. Ref. \onlinecite{optical_conductivity_2dQCP_Sachdev,optical_conductivity_2dQCP_Chubukov_Maslov}. 

Beyond the clean limit, we must consider the effects of a finite scattering rate for the fermions (due to the usual Fermi liquid interaction terms as well as the scattering off of pseudo-Goldstone bosons). Extracting the temperature scaling of the DC conductivity will likely require a consideration of spatial disorder as well\cite{sachdev_criticalFS_1,sachdev_criticalFS_2}. It has also been noted that in similar models of non-Fermi liquids, $\omega/T$ scaling may break down at finite temperature due to a 'thermal' contribution to the fermion self-energy \cite{torroba_thermal_NFL}, so it is a matter of interest to more carefully delineate the range of frequencies where the optical conductivity scales as in Fig. \ref{fig:conductivity_crossover}. We leave such analysis to future work. 

\section{Anisotropic Systems}
\label{Sec:Anisotropy}

An interesting scenario to consider is a 2D system that conserves only the dipole momentum along one direction (say the dipole conservation is along $x$-direction), and in the other direction, the fermions can have a bare hopping term. Such kind of model is easily realizable in the tilted optical lattice experiments. In this situation, after spontaneous symmetry breaking of the dipolar symmetry, a generic low-energy theory consists of a fermi surface coupled only to one Goldstone mode $\varphi^x$. The one-loop self-energy and propagator for $\varphi^x$ may simply be read off from the $\hat{x}\hat{x}$-component of Eq. \ref{eq:OneLoopSelfEnergy}:
\begin{equation}\label{eq:OneLoopSelfEnergyAniso}
\begin{split}
   &\Pi^{xx}(i\omega,\vec{p}) \approx -\gamma \frac{\abs{\omega}}{\abs{\vec{p}}}\sin^2\theta_p\\
   &D(i\omega,\vec{p})= \frac{g^2}{\omega^2+v^2_1\abs{p}^2+v^2_2p_x^2 + \gamma g^2\abs{\omega/\vec{p}}\sin^2\theta_p}
   \end{split}
\end{equation}
where $\theta_p=\arctan(p_y/p_x)$ and $\gamma$ is the same as in Eq. \ref{eq:OneLoopSelfEnergy}. The corresponding fermion self-energy is computed in Appendix \ref{Sec:AppFermionSE} for $\vec{q}$ on the fermi surface, and is given by

\begin{equation} \label{eq:FermionSelfEnergyAppendixAnisotropicMainText}
\begin{split}
\Sigma^{T}(i\omega,\vec{q}) &= -i\sign(\omega)\biggr(\frac{1}{6\sqrt{3}\pi^2}\frac{g^4v_F^2\kappa\cos^4\theta_F}{N(v^2_1+v^2_2\sin^2\theta_F)^2}\biggr)^{\frac{1}{3}}\abs{\omega}^{\frac{2}{3}}\\
&=-i\sign(\omega)(E_{NFL}(\vec{q}))^{1/3}\abs{\omega}^{2/3}
\end{split}
\end{equation}
where $\theta_F$ is the angle that the normal vector to the fermi surface at $\vec{q}$ makes with the $x$-axis. Notably, the NFL energy scale depends strongly on the position on the fermi surface, vanishing when $\theta_F = \pm \pi/2$. Thus, Fermi liquid behavior is present in the vicinity of these 'cold spots'. Such a $\cos^4\theta_F$ angular dependence of the NFL energy scale is similar to the behavior of the model of spin-orbit coupled ferromagnets which break rotational and spin symmetries \cite{bahri_potter_goldstone_NFL}. We note that the 3-point coupling of $\varphi_x$ to the fermi surface has an identical angular dependence as in this ferromagnetic model. The coefficient $v^2_2$ in the Goldstone action enters into the formula for $E_{NFL}(\vec{q})$ and contributes additional angular dependence (along with the intrinsic angular dependence of $v_F, \kappa$). 

\begin{figure}
    \centering
    \includegraphics[width=0.6\linewidth]{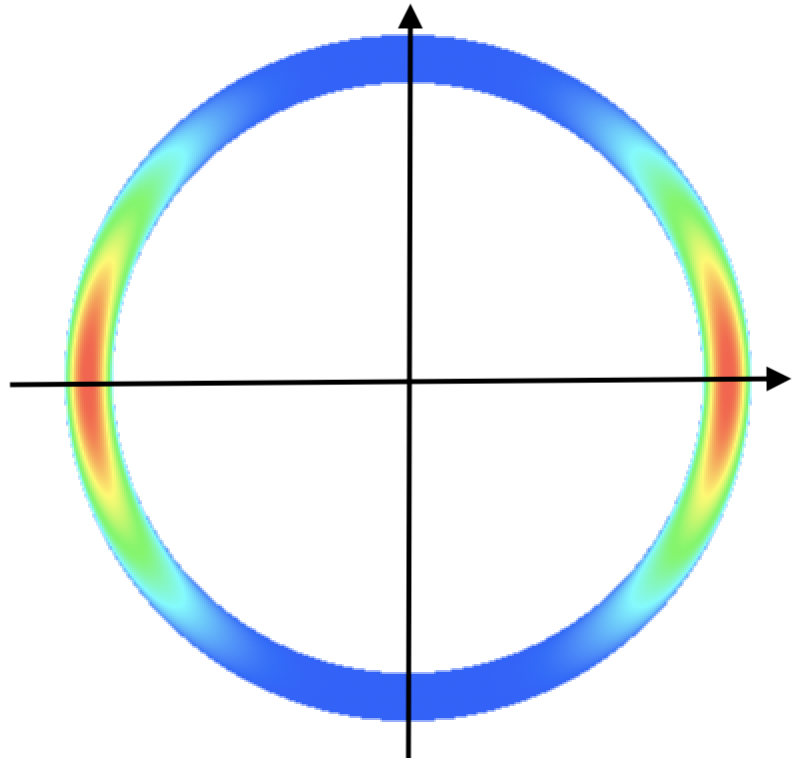}
    \caption{Depiction of a circular Fermi surface in a theory that conserves only one component of the dipole moment; the NFL energy scale is largest at the hot spots at $k_y=0$ and vanishes at the cold spots $k_x=0$. Color indicates $E_{NFL}\sim \cos^4\theta_F$.}
    \label{fig:cold_spots}
\end{figure}

The presence of these cold spots on the Fermi surface, sketched in Fig. \ref{fig:cold_spots} can be understood by recalling that the Goldstone mode at momentum $\vec{p}$ is most relevantly coupled to the parts of the Fermi surface parallel to $\vec{p}$. When $\theta_F = 0, \pi$, the most relevant Goldstone mode has perfect overlap with the transverse part of the Goldstone mode in Eq. \ref{eq:NGBPropagatorMainText}, which is the only part responsible for the singular fermion self-energy. On the other hand, when $\theta_F=\pm \pi/2$, the relevant Goldstone mode is purely longitudinal, which does not couple strongly to the fermi surface.

We can compute the optical conductivity similarly to the fully dipolar-symmetric case by exploiting the fact that the Goldstone $\varphi^x$ couples like the $x$-component of the electromagnetic vector potential. Likewise defining $K_{ab}(\omega,\vec{q})$ as the quadratic piece of the $\vec{A}$ effective action and $K^0_{ab}$ as the free-fermion contribution, and recognizing that $\Pi(\vec{q},\omega) = K_{xx}(\vec{q},\omega)$, we have
\begin{equation}
    K_{ab}(0,\omega)  = K_{ab}^0(0,\omega) + g^2K_{ac}^0(0,\omega)\frac{\delta_{cx}\delta_{dx}}{\omega^2-g^2K^0_{xx}}K^0_{db}(0,\omega)
\end{equation}

In the limit of $\omega\to 0$, we can see that $K_{xx}=K_{xy}= 0$, and thus the only nonvanishing part of the optical conductivity is
\begin{equation}\label{eq:anisotropic_conductivity}
\sigma_{yy}(0,\omega) \approx \frac{i}{\omega}\biggr(K^0_{yy} - \frac{(K^0_{xy})^2}{K^0_{xx}}\biggr).
\end{equation}
Thus, the uniform current along the $x$-direction will always vanish (as expected in a system that conserves the $x$-component of the dipole moment), and the conductivity along the $y$-direction is renormalized by a constant that only depends on the properties of the Fermi surface and is independent of the coupling strength. We note that this expression for the optical conductivity is also rendered exact upon passing to an IR patch theory\cite{GiftsFromAnomalies}.

\section{Conclusions and Discussions}
\label{Sec:conclusion}
In this work, we have shown that spontaneously breaking dipole symmetries in fermionic systems leads to compressible states with a variety of universal, exotic properties. In 1D, we showed that the dipole order parameter exhibits long-range order and that single fermions are confined with a drastic suppression of the tunneling density of states, results which can be understood via the emergent anomaly of a low-energy Dirac theory or a duality with the Bose-Einstein insulator theory\cite{LakeDBHM, Lake_Tilted_Chain, zechmann2022fractonic}. In 2D, the low-energy physics is identical to that of the famous non-Fermi Liquid consisting of a Fermi surface coupled to a dynamical $U(1)$ gauge field; based on the properties of the associated RG fixed point, we were able to predict an NFL/FL finite-temperature crossover controlled by the slight breaking of dipole symmetry. Furthermore, such explicit dipole symmetry breaking leads to a nonzero ``Drude" contribution to the optical conductivity which scales as $T^{-2/3}$ above a crossover temperature. 

We have restricted our analysis to systems that only break the dipolar symmetry. As suggested by our mean-field results in \ref{Sec:MF}, various types of spin order may also be present in even simple models of dipole-conserving fermions. The low-momentum coupling of the Fermi surface to such Goldstone modes resulting from internal symmetry breaking generically carries an extra factor of the momentum $\vec{p}$ compared to the dipolar Goldstones\cite{CriterionVishwanath}; hence, we do not expect them to modify the low-energy NFL physics. However, the presence of multiple non-overlapping fermi surfaces will modify the details of the patch theory beyond what we discussed in this work. Furthermore, quantum critical fluctuations of magnetic order parameters could lead to further modifications of the non-Fermi Liquid properties of this system beyond what is achieved by the dipolar Goldstones. Spin-orbit coupled Hamiltonians may be possible to realize as a mean-field ground state of a more complicated dipole-conserving model, and in principle the resulting quantum geometry from the Bloch states can affect the properties of the dipolar Goldstones. We leave such investigations to future work. 


\section*{Acknowledgements}
We thank Ethan Lake and Ribhu Kaul for helpful discussions. AA and ZB are supported by startup funds from the Pennsylvania State University.

\textit{Note: During the completion of this work we learned of a related work by E. Lake and T. Senthil \cite{lakeDFHM}.}

\bibliographystyle{unsrt}
\bibliography{arxivsubmission}

\begin{thebibliography}{10}

\bibitem{mcgreevy_generalized_symmetries}
John McGreevy.
\newblock Generalized symmetries in condensed matter.
\newblock {\em Annual Review of Condensed Matter Physics}, 14(1):57--82, 2023.

\bibitem{Gromov_Multipole}
Andrey Gromov.
\newblock Towards classification of fracton phases: The multipole algebra.
\newblock {\em Phys. Rev. X}, 9:031035, Aug 2019.

\bibitem{fractonreview1}
Rahul~M. Nandkishore and Michael Hermele.
\newblock Fractons.
\newblock {\em Annual Review of Condensed Matter Physics}, 10(1):295--313,
  2019.

\bibitem{fractonreview2}
Michael Pretko, Xie Chen, and Yizhi You.
\newblock Fracton phases of matter.
\newblock {\em International Journal of Modern Physics A}, 35(06):2030003,
  2020.

\bibitem{bulmash2023multipole}
Daniel Bulmash, Oliver Hart, and Rahul Nandkishore.
\newblock Multipole groups and fracton phenomena on arbitrary crystalline
  lattices, 2023.

\bibitem{pai_pretko_nandkishore_fractonic_random_circuits}
Shriya Pai, Michael Pretko, and Rahul~M. Nandkishore.
\newblock Localization in fractonic random circuits.
\newblock {\em Phys. Rev. X}, 9:021003, Apr 2019.

\bibitem{sala2020_ergodicity_from_dipole_cons}
Pablo Sala, Tibor Rakovszky, Ruben Verresen, Michael Knap, and Frank Pollmann.
\newblock Ergodicity breaking arising from hilbert space fragmentation in
  dipole-conserving hamiltonians.
\newblock {\em Physical Review X}, 10(1):011047, 2020.

\bibitem{aidelsburger_fragmentation_tilted}
Thomas Kohlert, Sebastian Scherg, Pablo Sala, Frank Pollmann, Bharath
  Hebbe~Madhusudhana, Immanuel Bloch, and Monika Aidelsburger.
\newblock Exploring the regime of fragmentation in strongly tilted
  fermi-hubbard chains.
\newblock {\em Phys. Rev. Lett.}, 130:010201, Jan 2023.

\bibitem{fractonhydro1}
Andrey Gromov, Andrew Lucas, and Rahul~M. Nandkishore.
\newblock Fracton hydrodynamics.
\newblock {\em Phys. Rev. Res.}, 2:033124, Jul 2020.

\bibitem{fractonhydro2}
Paolo Glorioso, Luca~V. Delacrétaz, Xiao Chen, Rahul~M. Nandkishore, and
  Andrew Lucas.
\newblock {Hydrodynamics in lattice models with continuous non-Abelian
  symmetries}.
\newblock {\em SciPost Phys.}, 10:015, 2021.

\bibitem{fractonhydro3}
Paolo Glorioso, Jinkang Guo, Joaquin~F. Rodriguez-Nieva, and Andrew Lucas.
\newblock Breakdown of hydrodynamics below four dimensions in a fracton fluid.
\newblock {\em Nature Physics}, 18(8):912--917, 2022.

\bibitem{subdiffusion1}
Jason Iaconis, Sagar Vijay, and Rahul Nandkishore.
\newblock Anomalous subdiffusion from subsystem symmetries.
\newblock {\em Phys. Rev. B}, 100:214301, Dec 2019.

\bibitem{subdiffusion2}
Johannes Feldmeier, Pablo Sala, Giuseppe De~Tomasi, Frank Pollmann, and Michael
  Knap.
\newblock Anomalous diffusion in dipole- and higher-moment-conserving systems.
\newblock {\em Phys. Rev. Lett.}, 125:245303, Dec 2020.

\bibitem{subdiffusion3}
Jason Iaconis, Andrew Lucas, and Rahul Nandkishore.
\newblock Multipole conservation laws and subdiffusion in any dimension.
\newblock {\em Phys. Rev. E}, 103:022142, Feb 2021.

\bibitem{freezing}
Alan Morningstar, Vedika Khemani, and David~A. Huse.
\newblock Kinetically constrained freezing transition in a dipole-conserving
  system.
\newblock {\em Phys. Rev. B}, 101:214205, Jun 2020.

\bibitem{khemani_hermele_nandkishore_tilted}
Vedika Khemani, Michael Hermele, and Rahul Nandkishore.
\newblock Localization from hilbert space shattering: From theory to physical
  realizations.
\newblock {\em Phys. Rev. B}, 101:174204, May 2020.

\bibitem{huse_bakr_tilted_lattice}
Elmer Guardado-Sanchez, Alan Morningstar, Benjamin~M. Spar, Peter~T. Brown,
  David~A. Huse, and Waseem~S. Bakr.
\newblock Subdiffusion and heat transport in a tilted two-dimensional
  fermi-hubbard system.
\newblock {\em Phys. Rev. X}, 10:011042, Feb 2020.

\bibitem{scherg_aidelsburger_2021observing}
Sebastian Scherg, Thomas Kohlert, Pablo Sala, Frank Pollmann, Bharath
  Hebbe~Madhusudhana, Immanuel Bloch, and Monika Aidelsburger.
\newblock Observing non-ergodicity due to kinetic constraints in tilted
  fermi-hubbard chains.
\newblock {\em Nature Communications}, 12(1):4490, 2021.

\bibitem{PengYe1}
Jian-Keng Yuan, Shuai~A. Chen, and Peng Ye.
\newblock Fractonic superfluids.
\newblock {\em Phys. Rev. Res.}, 2:023267, Jun 2020.

\bibitem{PengYe2}
Shuai~A. Chen, Jian-Keng Yuan, and Peng Ye.
\newblock Fractonic superfluids. ii. condensing subdimensional particles.
\newblock {\em Phys. Rev. Res.}, 3:013226, Mar 2021.

\bibitem{MultipolarMerminWagner}
Charles Stahl, Ethan Lake, and Rahul Nandkishore.
\newblock Spontaneous breaking of multipole symmetries.
\newblock {\em Phys. Rev. B}, 105:155107, Apr 2022.

\bibitem{kapustin_dipole}
Anton Kapustin and Lev Spodyneiko.
\newblock Hohenberg-mermin-wagner-type theorems and dipole symmetry.
\newblock {\em Phys. Rev. B}, 106:245125, Dec 2022.

\bibitem{LakeDBHM}
Ethan Lake, Michael Hermele, and T.~Senthil.
\newblock Dipolar bose-hubbard model.
\newblock {\em Phys. Rev. B}, 106:064511, Aug 2022.

\bibitem{Lucas_Dipolar_Goldstone_HD}
Paolo Glorioso, Xiaoyang Huang, Jinkang Guo, Joaquin~F Rodriguez-Nieva, and
  Andrew Lucas.
\newblock Goldstone bosons and fluctuating hydrodynamics with dipole and
  momentum conservation.
\newblock {\em Journal of High Energy Physics}, 2023(5):1--43, 2023.

\bibitem{stahl2023fracton}
Charles Stahl, Marvin Qi, Paolo Glorioso, Andrew Lucas, and Rahul Nandkishore.
\newblock Fracton superfluid hydrodynamics.
\newblock {\em arXiv preprint arXiv:2303.09573}, 2023.

\bibitem{HLR}
B.~I. Halperin, Patrick~A. Lee, and Nicholas Read.
\newblock Theory of the half-filled landau level.
\newblock {\em Phys. Rev. B}, 47:7312--7343, Mar 1993.

\bibitem{CriterionVishwanath}
Haruki Watanabe and Ashvin Vishwanath.
\newblock Criterion for stability of goldstone modes and fermi liquid behavior
  in a metal with broken symmetry.
\newblock {\em Proceedings of the National Academy of Sciences},
  111(46):16314--16318, 2014.

\bibitem{dopemott}
Patrick~A. Lee, Naoto Nagaosa, and Xiao-Gang Wen.
\newblock Doping a mott insulator: Physics of high-temperature
  superconductivity.
\newblock {\em Rev. Mod. Phys.}, 78:17--85, Jan 2006.

\bibitem{lohneysen_HMM_theory}
Hilbert~v L{\"o}hneysen, Achim Rosch, Matthias Vojta, and Peter W{\"o}lfle.
\newblock Fermi-liquid instabilities at magnetic quantum phase transitions.
\newblock {\em Reviews of Modern Physics}, 79(3):1015, 2007.

\bibitem{senthil_spinon_FS}
T.~Senthil.
\newblock Theory of a continuous mott transition in two dimensions.
\newblock {\em Phys. Rev. B}, 78:045109, Jul 2008.

\bibitem{mross_mcgreevy_liu_senthil_NFL_eps_N}
David~F. Mross, John McGreevy, Hong Liu, and T.~Senthil.
\newblock Controlled expansion for certain non-fermi-liquid metals.
\newblock {\em Phys. Rev. B}, 82:045121, Jul 2010.

\bibitem{sslee2009}
Sung-Sik Lee.
\newblock Low-energy effective theory of fermi surface coupled with u(1) gauge
  field in $2+1$ dimensions.
\newblock {\em Phys. Rev. B}, 80:165102, Oct 2009.

\bibitem{sachdevQPT}
Subir Sachdev.
\newblock {\em Quantum Phase Transitions}.
\newblock Cambridge University Press, 2 edition, 2011.

\bibitem{metlitski_sachdev_ising}
Max~A. Metlitski and Subir Sachdev.
\newblock Quantum phase transitions of metals in two spatial dimensions. i.
  ising-nematic order.
\newblock {\em Phys. Rev. B}, 82:075127, Aug 2010.

\bibitem{mandal_transverse_gauge_field}
Ipsita Mandal.
\newblock Critical fermi surfaces in generic dimensions arising from transverse
  gauge field interactions.
\newblock {\em Phys. Rev. Res.}, 2:043277, Nov 2020.

\bibitem{sachdev_criticalFS_1}
Ilya Esterlis, Haoyu Guo, Aavishkar~A. Patel, and Subir Sachdev.
\newblock Large-$n$ theory of critical fermi surfaces.
\newblock {\em Phys. Rev. B}, 103:235129, Jun 2021.

\bibitem{sachdev_criticalFS_2}
Haoyu Guo, Aavishkar~A. Patel, Ilya Esterlis, and Subir Sachdev.
\newblock Large-$n$ theory of critical fermi surfaces. ii. conductivity.
\newblock {\em Phys. Rev. B}, 106:115151, Sep 2022.

\bibitem{Schwinger_Model_OG}
Julian Schwinger.
\newblock Gauge invariance and mass. ii.
\newblock {\em Phys. Rev.}, 128:2425--2429, Dec 1962.

\bibitem{GiftsFromAnomalies}
Zhengyan {Darius Shi}, Hart {Goldman}, Dominic~V. {Else}, and T.~{Senthil}.
\newblock {Gifts from anomalies: Exact results for Landau phase transitions in
  metals}.
\newblock {\em arXiv e-prints}, page arXiv:2204.07585, April 2022.

\bibitem{Peskin}
Michael~Edward Peskin and Daniel~V. Schroeder.
\newblock {\em {An Introduction to Quantum Field Theory}}.
\newblock Westview Press, 1995.
\newblock Reading, USA: Addison-Wesley (1995) 842 p.

\bibitem{kogut_sinclair}
J.~Kogut and D.~K. Sinclair.
\newblock Quark confinement and the evasion of goldstone's theorem in 1+1
  dimensions.
\newblock {\em Phys. Rev. D}, 12:1742--1753, Sep 1975.

\bibitem{kogut_susskind}
J.~Kogut and Leonard Susskind.
\newblock How quark confinement solves the
  $\ensuremath{\eta}\ensuremath{\rightarrow}3\ensuremath{\pi}$ problem.
\newblock {\em Phys. Rev. D}, 11:3594--3610, Jun 1975.

\bibitem{casher_kogut_susskind}
A.~Casher, J.~Kogut, and Leonard Susskind.
\newblock Vacuum polarization and the absence of free quarks.
\newblock {\em Phys. Rev. D}, 10:732--745, Jul 1974.

\bibitem{fradkin_2013}
Eduardo Fradkin.
\newblock {\em Field Theories of Condensed Matter Physics}.
\newblock Cambridge University Press, 2 edition, 2013.

\bibitem{Note1}
Surprisingly, the spatial current $j_1$ is a dipole-invariant operator; this
  follows from the perfectly linear dispersion of the Dirac fermion. Quadratic
  and higher corrections are proportional to 2nd and higher derivatives of the
  dispersion near $k_F$, and are thus irrelevant perturbations to the Luttinger
  liquid action. One may ask about the interaction terms involving $\varphi $
  generated by such corrections to the dispersion, but upon noticing that the
  coupling $g$ becomes a mass scale, $\varphi $ also acquires dimensions of
  mass, and thus all such interaction terms are rendered irrelevant.

\bibitem{Lake_Tilted_Chain}
Ethan Lake, Hyun-Yong Lee, Jung~Hoon Han, and T.~Senthil.
\newblock Dipole condensates in tilted bose-hubbard chains.
\newblock {\em Phys. Rev. B}, 107:195132, May 2023.

\bibitem{zechmann2022fractonic}
Philip Zechmann, Ehud Altman, Michael Knap, and Johannes Feldmeier.
\newblock Fractonic luttinger liquids and supersolids in a constrained
  bose-hubbard model.
\newblock {\em Phys. Rev. B}, 107:195131, May 2023.

\bibitem{pretko_1d_confinement_fractons}
Shriya Pai and Michael Pretko.
\newblock Fractons from confinement in one dimension.
\newblock {\em Phys. Rev. Res.}, 2:013094, Jan 2020.

\bibitem{Watanabe_Goldstone_Theorem}
Haruki Watanabe and Hitoshi Murayama.
\newblock Unified description of nambu-goldstone bosons without lorentz
  invariance.
\newblock {\em Phys. Rev. Lett.}, 108:251602, Jun 2012.

\bibitem{Max_Cooper_Pairing_NFL}
Max~A. Metlitski, David~F. Mross, Subir Sachdev, and T.~Senthil.
\newblock Cooper pairing in non-fermi liquids.
\newblock {\em Phys. Rev. B}, 91:115111, Mar 2015.

\bibitem{Note2}
Strictly speaking, tuning $v_2^2=-v_1^2$ in the free Goldstone action would
  also lead to a gauge-theoretic structure by making the action invariant under
  $\varphi ^a\protect \mapsto \varphi ^a-\partial ^a\chi (x)$. However, this is
  an extremely fine-tuned condition.

\bibitem{oganesyan_kivelson_fradkin_goldstone_NFL}
Vadim Oganesyan, Steven~A. Kivelson, and Eduardo Fradkin.
\newblock Quantum theory of a nematic fermi fluid.
\newblock {\em Phys. Rev. B}, 64:195109, Oct 2001.

\bibitem{ruhman_berg_goldstone_NFL}
Jonathan Ruhman and Erez Berg.
\newblock Ferromagnetic and nematic non-fermi liquids in spin-orbit-coupled
  two-dimensional fermi gases.
\newblock {\em Phys. Rev. B}, 90:235119, Dec 2014.

\bibitem{bahri_potter_goldstone_NFL}
Yasaman Bahri and Andrew~C. Potter.
\newblock Stable non-fermi-liquid phase of itinerant spin-orbit coupled
  ferromagnets.
\newblock {\em Phys. Rev. B}, 92:035131, Jul 2015.

\bibitem{mandal_boson_mass_NFL_crossover}
Ipsita Mandal and Rafael~M. Fernandes.
\newblock Valley-polarized nematic order in twisted moir\'e systems: In-plane
  orbital magnetism and crossover from non-fermi liquid to fermi liquid.
\newblock {\em Phys. Rev. B}, 107:125142, Mar 2023.

\bibitem{nayak_wilczek_nFL_fixed_point}
Chetan Nayak and Frank Wilczek.
\newblock Non-fermi liquid fixed point in 2 +1 dimensions.
\newblock {\em Nuclear Physics B}, 417(3):359--373, apr 1994.

\bibitem{nayak_wilczek_nFLRG}
Chetan Nayak and Frank Wilczek.
\newblock Renormalization group approach to low temperature properties of a
  non-fermi liquid metal.
\newblock {\em Nuclear Physics B}, 430(3):534--562, 1994.

\bibitem{optical_conductivity_2dQCP_Sachdev}
Andreas Eberlein, Ipsita Mandal, and Subir Sachdev.
\newblock Hyperscaling violation at the ising-nematic quantum critical point in
  two-dimensional metals.
\newblock {\em Phys. Rev. B}, 94:045133, Jul 2016.

\bibitem{ersatzFL}
Dominic~V. Else, Ryan Thorngren, and T.~Senthil.
\newblock Non-fermi liquids as ersatz fermi liquids: General constraints on
  compressible metals.
\newblock {\em Phys. Rev. X}, 11:021005, Apr 2021.

\bibitem{shi2022loop}
Zhengyan~Darius Shi, Dominic~V Else, Hart Goldman, et~al.
\newblock Loop current fluctuations and quantum critical transport.
\newblock {\em arXiv preprint arXiv:2208.04328}, 2022.

\bibitem{optical_conductivity_2dQCP_Chubukov_Maslov}
Andrey~V. Chubukov and Dmitrii~L. Maslov.
\newblock Optical conductivity of a two-dimensional metal near a quantum
  critical point: The status of the extended drude formula.
\newblock {\em Phys. Rev. B}, 96:205136, Nov 2017.

\bibitem{torroba_thermal_NFL}
Huajia Wang and Gonzalo Torroba.
\newblock Non-fermi liquids at finite temperature: Normal-state and infrared
  singularities.
\newblock {\em Phys. Rev. B}, 96:144508, Oct 2017.

\bibitem{lakeDFHM}
Ethan Lake and T.~Senthil.
\newblock Non-fermi liquids from kinetic constraints in tilted optical
  lattices.
\newblock {\em Phys. Rev. Lett.}, 131:043403, Jul 2023.

\bibitem{pa_lee_gauge_field}
Patrick~A. Lee.
\newblock Gauge field, aharonov-bohm flux, and high-${T}_{c}$
  superconductivity.
\newblock {\em Phys. Rev. Lett.}, 63:680--683, Aug 1989.

\bibitem{Note3}
We have not dropped the term $\protect \mathaccentV {bar}016{\psi }_\theta
  \partial _t \psi _\theta $ even though it is irrelevant under $z=3$ scaling,
  because doing so would kill any dynamics in the action \ref
  {eq:FinalPatchActionApp}. Including this irrelevant term ensures that the
  frequency contour integrals in the one-loop diagrams reproduce the correct
  results \cite {sslee2009}.

\end{thebibliography}

\appendix
\onecolumngrid
\section{Mean-field theory}
\label{Sec:AppMFTDetails}
Here, we start from Eq. \ref{eq:MicroHamiltonian} with $U=0$; the usual Hartree-Fock procedure with the ansatz $\avg{c^\dag_{\bf{i}+\bf{R},s}c_{\bf{i},s'}} = R^{\bf{R}}\delta_{ss'} + D^{\bf{R}}\sigma^z_{ss'}$ (without loss of generality assuming polarization along the z-axis) yields a mean-field Hamiltonian of the form

\begin{equation} \label{eq:HartreeFockHamiGeneral}
\begin{split}
\hat{H}_{MF} &= \sum_{i,j,\mathbf{R},s}\mathcal{A}_{ij}^{\bf{R}}\biggr(2R^{\bf{R}}c^\dag_{is}c_{i+\bf{R},s}-R^{j-i}c^\dag_{is}c_{j,s} + h.c. \biggr) - \sum_{i,j,\bf{R},s,s'}\mathcal{A}_{ij}^{\bf{R}}\sigma^z_{ss'}\biggr(D^{j-i}c^\dag_{is}c_{j,s'} + h.c.\biggr)\\
&-\sum_{i,j,\mathbf{R}}\mathcal{A}_{ij}^{\mathbf{R}}\biggr(-2\abs{R^{j-i}}^2-2\abs{D^{j-i}}^2 + 4\abs{R^{\bf{R}}}^2\biggr)
\end{split}
\end{equation}
We choose the hopping matrix to be of the nearest-neighbor form,
\begin{equation}\label{eq:LakeHopping}
    \mathcal{A}^{a}_{ij} = \sum_b\biggr(-t\delta_{ab}(\delta_{i,j+a}+\delta_{i,j-a}) - t'(1-\delta_{ab})(\delta_{i,j+b}+\delta_{i,j-b})\biggr)
\end{equation}
then the mean-field Hamiltonian takes the following simple form:
\begin{equation} \label{eq:HartreeFockHami}
\begin{split}
    \hat{H}_{MF} &= -2\tilde{t}\sum_{i,a}\biggr(R^a_{\uparrow}c^\dag_{i,\uparrow}c_{i+a,\uparrow}+R^a_{\downarrow}c^\dag_{i,\downarrow}c_{i+a,\downarrow} + h.c.\biggr) +\tilde{t}\sum_{i,a}\biggr(\abs{R^a_{\uparrow}+R^a_{\downarrow}}^2 - \abs{R^a_{\uparrow}-R^a_{\downarrow}}^2\biggr),
\end{split}
\end{equation}
where $\tilde{t}=t+(d-1)t'$ and $\avg{c^\dag_{{i}+a,\uparrow(\downarrow)}c_{{i},\uparrow(\downarrow)}} = R^a_{\uparrow(\downarrow)}$. The mean-field consistency condition is given by
\begin{equation}\label{eq:SelfConsistency}
    \sum_i\avg{c^\dag_{\bf{i}+\bf{R},s}c_{\bf{i},s'}}=  \mathcal{V}\int\frac{d^dk}{(2\pi)^2}e^{-i\vec{k}\cdot\mathbf{R}}g_{ss'}(\vec{k})
\end{equation}
where $g_{ss'}(\vec{k})= \avg{c^\dag_{\vec{k},s}c_{\vec{k},s'}}$.

In the simple case of Eq. \ref{eq:HartreeFockHami}, $g_{ss'}(\vec{k})$ is diagonal in the spin indices, with the entries just given by the usual Fermi sea distribution functions $n_F(\epsilon_s(\vec{k})-\mu)$. (Immediately, we observe that if the mean-field Hamiltonian of Eq. \ref{eq:HartreeFockHami} is an insulator, it is a trivial flat-band insulator. For more complicated models, $g_{ss'}(\vec{k})$ can have a more complicated momentum dependence and thus nontrivial band insulators can in principle emerge as mean-field ground states.) In the following, we work at generic filling and take $T=0$ for simplicity. 

The self-consistency condition \ref{eq:SelfConsistency} allows two symmetry breaking patterns $(R^x,R^y)=(R_{1D},0)$ (breaking only one component of the dipole symmetry) and $R^x=R^y=R_{2D}$ (breaking both components). We identify the mean-field ground state by minimizing the ground state energy of $\hat{H}_{MF}$, crucially including the condensation energy. First, we note that when $\tilde{t}>0$, the condensation energy term forces $R^{a}_{\uparrow} = -R^a_{\downarrow}$, so that the Fermi seas for each spin species are separated by a $(\pi,\pi)$ momentum (and for $\tilde{t}<0$ the Fermi seas are perfectly degenerate). Using the self-consistency condition \ref{eq:SelfConsistency}, we may write down the mean-field energy entirely in terms of $R^a_{\uparrow/\downarrow}$, assuming them to be real numbers whose relative sign is chosen to minimize the condensation energy:

\begin{equation}\label{eq:MeanFieldEnergy}
    \frac{E_{MF}}{\mathcal{V}}=  -4p_{SSB}\abs{\tilde{t}}\biggr(\abs{R_{\uparrow}(n_\uparrow)}^2+\abs{R_{\downarrow}(n_\downarrow)}^2 +\abs{R^a_{\uparrow}(n_\uparrow)R^a_{\downarrow}(n_\downarrow)} \biggr)
\end{equation}

Here, $p_{SSB}$ refers to the number of components of the dipole symmetry that are spontaneously broken, and the quantities $R_{\uparrow(\downarrow)}$ are given by
\begin{equation}
    R_{\uparrow(\downarrow)}(n_{\uparrow(\downarrow)}) = \int_{FS_{\uparrow(\downarrow)}}\frac{d^dk}{(2\pi)^d}\cos(k_x)
\end{equation}
where the integration region is a Fermi sea fixed by the population $n_{\uparrow(\downarrow)}$ and the functional form of the dispersion relation $\epsilon(\vec{k})\propto -\sum_a\cos(k_a)$, with the sum $\sum_a$ running over the broken directions (here we have assumed that one of the broken directions is always $x$). This is a function that is easy to evaluate numerically, and all that is left to do is minimize $E_{MF}$ with respect to $n_{\uparrow}$, with the only relevant parameters to vary being $n$ (the number of electrons per site) and the number of broken components of the dipole symmetry. The following results refer to a 2-D system, though the computations may be done in any dimension. 

We thus find that for $0.6<n<1$, the system prefers to spontaneously break only one component of the dipole symmetry, leading to a quasi-1D state consisting of a Fermi sea with dispersion only along the direction of the broken dipole symmetry. For fillings $0.17<n<0.6$ electrons per site, both components of the dipole symmetry are spontaneously broken leading to two identical 2D Fermi seas; when $n<0.17$, the ground state is actually ferromagnetic as it becomes energetically favourable for all the fermions to settle into one spin species (i.e. $R^a_{\uparrow}>0$, $R^a_{\downarrow}=0$). The ground states at fillings $n, 2-n$ are identical by particle-hole symmetry.




The above conclusions are not universal; for more complicated hopping matrices $\mathcal{A}_{ij}$, the mean-field consistency condition will need to be checked for multiple bilinears $\rho^{\mathbf{R}}_{i}$, and more complicated dispersion relations may emerge (with the possibility of more exotic spin ordering as well). The nature of filling-tuned transitions between symmetry-breaking patterns is also model-dependent.

We may extract the fluctuations beyond mean-field theory by granting position-dependence to our ansatz: $\avg{c^\dag_{\bf{i}+\bf{R},s}c_{\bf{i},s'}} =R^{\bf{R}}(\mathbf{i})\delta_{ss'}$ (such an ansatz describes the case studied in most of this work). Accordingly, for a hopping matrix $\mathcal{A}^{\mathbf{R}}_{ij}$ as in Eq. \ref{eq:LakeHopping}, the condensation energy takes the form
\begin{equation}
    \begin{split}
        H_{condensate} &= \sum_{ij,a}\mathcal{A}_{ij}^a\biggr(2R^{{j-i}}(i+a)(R^{j-i}(i))^* - 4(R^{a}(i))^*R^{a}(j)\biggr)\\
        &= 2(t-t')\sum_{ia}\biggr((R^a(i+a))^*R^a(i)+(R^a(i-a))^*R^a(i)\biggr)-2t'\sum_{iab}\biggr((R^a(i+b))^*R^a(i)+(R^a(i-b))^*R^a(i)\biggr)
    \end{split}
\end{equation}

Focusing on the terms involving derivatives of the order parameter,
\begin{equation}
    H\sim \int d^dx\,\abs{a}^{2-d}\biggr(2t'\sum_{ab}\abs{\partial_aR^b}^2+2(t-t')\sum_a\abs{\partial_aR^a}^2 \biggr)
\end{equation}
where $a$ is the lattice constant. Following the parametrization $R^{a}(\mathbf{i}) = \abs{R}e^{i\varphi^a(\mathbf{i})}$ (assuming both components of the dipole order parameter condense), the coefficients of terms involving spatial derivatives of $\varphi^a$ can be directly related to the parameters $v^2_1, v^2_2, v_3^2, g$ appearing in the Goldstone action (Eq. \ref{eq:MicroscopicAction}). Indeed, in 2D, we have $v^2_1=4t'\abs{R}^2$ and $v^2_3=4(t-t')\abs{R}^2$. By reading off the Fermi-Goldstone interaction from the mean-field Hamiltonian, we see that the Fermi velocity $v_F\sim \tilde{t}\abs{R}a$ (up to an $O(1)$ filling-dependent constant), and $g=a^{-1}$.




\section{General form of Fermion-Goldstone coupling}
\label{Sec:AppGoldstoneCouplings}

Here, we justify the form of the theory in Eq. \ref{eq:MicroscopicAction} by using a result from Ref. \onlinecite{CriterionVishwanath}. We suppose the system Hamiltonian has a global $U(1)^{\times d}$ symmetry with the dipole charges $\hat{Q}_a$. Consider the ground state that spontaneously breaks all $d$ generators. Thus we may decompose the effective hamiltonian (after discarding the gapped Higgs modes) as
\begin{equation}
    \hat{H} =  \hat{H}_{0}(\bar{\psi},\psi) + \hat{H}_{NGB}(\varphi^a)+\hat{H}_{int}
\end{equation}
where $\hat{H}_0=\sum_i\hat{\mathcal{H}}_{0}(i)$ is the fermionic part of the mean-field Hamiltonian which breaks the symmetry generated by $\hat{Q}_a$, $\hat{H}_{NGB}$ involves only Goldstone fields, and $\hat{H}_{int}$ describes fermion-Goldstone interactions. The interaction term can be written as a series expansion in the Goldstone fields $\varphi^a$,
\begin{equation}
    \begin{split}
        \hat{H}_{int} &= \hat{H}_{int}^{(1)} + \hat{H}_{int}^{(2)}\ldots\\
        \hat{H}_{int}^{(1)} &=\sum_i \varphi_c(i)\frac{\delta\hat{H}^{(1)}_{int}}{\delta \varphi^c}\\
        \hat{H}_{int}^{(2)} &= \frac12\sum_i\varphi_a(i)\varphi_b(i)\frac{\delta^2\hat{H}^{(2)}_{int}}{\delta \varphi^a\delta \varphi^b}
    \end{split}
\end{equation}
where the functional derivatives do not depend on Goldstone operators. Demanding that the total Hamiltonian $\hat{H}$ is symmetric, we have 
\begin{equation}\label{eq:GeneralGoldstoneFermi}
\begin{split}
\hat{H}_{int}^{(1)} &= -i\sum_i[\varphi_a(i)Q^a,\hat{\mathcal{H}}_{0}(i)]\\
\hat{H}_{int}^{(2)} &= -\frac12\sum_i[\varphi_a(i)Q^a,[\varphi_b(i)Q^b,\hat{\mathcal{H}}_{0}(i)]] = -\frac{i}{2}\sum_i[\varphi_b(i)Q^b, \mathcal{\hat{H}}_{int}^{(1)}(i)] \\ 
&\ldots \\
\hat{H}_{int}^{(n)} & = -\frac{i}{n}\sum_i[\varphi_{a_n}(i)Q^{a_n}, \mathcal{\hat{H}}_{int}^{(n-1)}(i)]
\end{split}
\end{equation}

Now, as is the case in our system, when the charge-$k$ operators $\rho_{i}^{a;(k)}$ condense, they will generate $k$'th nearest-neighbor hoppings for the fermions. Thus, the mean-field Hamiltonian will have the general form in real space:
\begin{equation}
    \hat{\mathcal{H}}_0 = \sum_i \hat{\mathcal{H}}_{0}(i)\ \ \mathrm{with}\ \ \ \hat{\mathcal{H}}_{0}(i) = \sum_{\bf{r_j}}\biggr(t_{j}^*\rho_i^{(\bf{r}_j)}+t_{j}(\rho_i^{(\bf{r}_j)})^\dag\biggr) + \hat{\mathcal{H}}_{onsite}
\end{equation}

We now wish to compute the commutator $[Q^a,\hat{\mathcal{H}}_{0}(i)]$. Using the expression for the dipole charge $\hat{Q}_a = \sum_{\bf{j}} \vec{a}\cdot \vec{r}_{\bf{j}}\hat{n}_{\bf{j}}$ and Eq. \ref{eq:GeneralGoldstoneFermi}, we can show
\begin{equation}
    \hat{H}_{int}^{(1)} =  \sum_{s,a}\int[dq][dk][dk'] \delta^d(q-(k'-k))\varphi^a_qc^\dag_{\vec{k},s}c_{\vec{k}',s}v_{k,k'}^{(3,a)},
\end{equation}
where $[dq] = \frac{d^dq}{(2\pi)^d}$, with the vertex factor 
\begin{equation}
    v_{k,k'}^{(3,a)}=\vec{a}\cdot\sum_{j}(t^*_j\grad_ke^{i\vec{r}_j\cdot\vec{k}}+t_j\grad_{k'}e^{-i\vec{r}_j\cdot\vec{k}'})
\end{equation}

We can also compute:
\begin{equation}
    \begin{split}
        \hat{H}_{int}^{(2)} &=  -\frac{1}{2}\sum_i[\varphi_b(i)Q^b, \varphi_a(i)\sum_j\vec{a}\cdot\vec{r}_j\biggr(t_{j}^*\rho_i^{(\bf{r}_j)}-t_{j}(\rho_i^{(\bf{r}_j)})^\dag\biggr)]\\
         &=  \frac12\sum_{s,a,b}\int[dq'][dq][dk][dk'] \delta^d(q+q'-(k'-k))\varphi^a_{q}\varphi^b_{q'}c^\dag_{\vec{k},s}c_{\vec{k}',s}\sum_{j}(t^*_j\partial_a\partial_be^{i\vec{r}_j\cdot\vec{k}}+t_j\partial'_a\partial'_be^{-i\vec{r}_j\cdot\vec{k}'})\\
    \end{split}
\end{equation}

In the above computations, the sums over $j$ count only \textit{one} of the two lattice vectors $\vec{r}_j$, $-\vec{r}_j$, as this is how one sums over the independent complex order parameters for the dipole condensate. Because of the delta function, we may write $k\to k-q/2$, $k'\to k+q/2$. Then, in the limit of low momentum transfer ($q\to 0$), 
\begin{equation}
    \begin{split}           v_{k,k'}^{(3,a)}&=i\sum_{j}\vec{a}\cdot\vec{r}_j(t^*_je^{i\vec{r}_j\cdot\vec{k}}-t_je^{-i\vec{r}_j\cdot\vec{k}})e^{-i\vec{r}_j\cdot\vec{q}/2} \approx \partial_a\epsilon_k \\
    v_{k,k'}^{(4,a,b)}&=-\frac12\sum_{j}(\vec{a}\cdot\vec{r}_j)(\vec{b}\cdot\vec{r}_j)(t^*_je^{i\vec{r}_j\cdot\vec{k}}+t_je^{-i\vec{r}_j\cdot\vec{k}})e^{-i\vec{r}_j\cdot\vec{q}/2} \approx \frac12\partial_a\partial_b\epsilon_k,
    \end{split}
\end{equation}
where $\partial_a=\vec{a}\cdot\grad_k$. These are precisely the low-momentum couplings predicted by the general formula in Ref. \onlinecite{CriterionVishwanath}, using the fact that the commutator of the dipole operator $\hat{Q}_a$ and the momentum operator $\hat{p}_b$ (in the single-particle Hilbert space) is a constant $i\delta_{ab}$. 

Generalizing the previous computations, if the mean-field Hamiltonian is an $N$-band system and the dipole symmetry acts trivially in the unit cell Hilbert space, we have the \textit{small momentum exchange} form of the Hamiltonian as
\begin{equation}
    \hat{H}_0 = \sum_{\vec{k},s,s'}c^\dag_{k,s}h_{s,s'}(\vec{k})c_{k,s'}
\end{equation}
\begin{equation}
    \begin{split}
        \hat{H}_{int}^{(1)} &=  \sum_{s,s',a}\int[dq][dk][dk'] \delta^d(q-(k'-k))\varphi^a_qc^\dag_{\vec{k},s}c_{\vec{k}',s'}\partial_ah_{s,s'}(\frac12(\vec{k}+\vec{k}'))\\
        \hat{H}_{int}^{(2)} &=  \frac12\sum_{s,a,b}\int[dq'][dq][dk][dk'] \delta^d(q+q'-(k'-k))\varphi^a_{q}\varphi^b_{q'}c^\dag_{\vec{k},s}c_{\vec{k}',s'}\times \partial_a\partial_bh_{ss'}(\frac12(\vec{k}+\vec{k}'))
    \end{split}
\end{equation}
In general, it is not difficult to guess that
\begin{equation}
    \hat{H}_{int}^{(n)} = \frac{1}{n!}\sum_{s,a_1\ldots a_n}[dq_1\ldots dq_ndkdk']\delta((k'-k)-\sum q_i)\phi^{a_1}_{q_1}\ldots\phi^{a_n}_{q_n}c^\dag_{\vec{k},s}c_{\vec{k}',s'}\times \partial_{a_1}\ldots\partial_{a_n}h_{ss'}(\frac12(\vec{k}+\vec{k}'))
\end{equation}

We comment that the Goldstone-Fermion coupling is exactly given by the Peierls substitution $\vec{k}\mapsto \vec{k}-\vec{\varphi}$ in the Bloch Hamiltonian. This is not particularly surprising; when the fermion bilinear condenses as $\avg{c^\dag_{i+a}c_i} = R^ae^{i\varphi^a}$, the Goldstone mode $\varphi^a$ is exactly a hopping phase. Given that the Bloch Hamiltonian $h_{s,s'}(\vec{k})$ can have eigenvectors with a complicated dependence on momentum, it is possible that the fermion-NGB interactions can be affected by the nontrivial quantum geometry, completely analogously to how the electromagnetic response of free fermions is affected by the band geometry.

\section{Schwinger-Dyson Equations and Exact Goldstone Boson Propagator in 1D}
\label{Sec:AppSchwingerDyson}
Here, we clarify the arguments reviewed in the main text for the exact solution of the 1-D boson propagator in the theory \ref{eq:DiracAction}. A simple Schwinger-Dyson equation one may write down is (\cite{Peskin})

$$-i\delta(x-x_0)\delta(t-t_0) = \bigavg{\varphi(x_0,t_0)\frac{\delta}{\delta\varphi(x,t)}\int d^4y\mathcal{L}} = \bigavg{\varphi(x_0,t_0)\biggr(-\frac{1}{g^2}(\partial_{t}^2-v_B^2\partial_{x}^2)\biggr)\varphi(x,t)}+\bigavg{v_F\varphi(x_0,t_0)j^1(x,t)}$$

where all of the above correlation functions are assumed to be time-ordered. Now, suppose $\mathcal{F}$ is a functional that is invariant under $U(1)_A$ (as defined in the main text). Then, the precise statement of the axial anomaly takes the form of an \textit{anomalous Ward Identity}:

$$\bigavg{\mathcal{F}(\partial_\mu j^\mu_5+\frac{1}{\pi}\partial_t\varphi)} = 0$$

Now, we let $\mathcal{F} = \varphi$ and operate on the insertion of $\partial_\mu j^\mu_5+\frac{1}{\pi}\partial_t\varphi$ with $\partial_t$. Using $\partial_\mu j^\mu_5 = \partial_t j^1+v_F\partial_x j^0$ and using the usual Ward identity for current conservation $\avg{\mathcal{F}\partial_\mu j^\mu} = 0$ (for a $U(1)_V$-invariant functional $\mathcal{F}$) to freely replace $\partial_t\partial_x j^0$ with $-v_F\partial_x^2 j^1$, we arrive at

$$\bigavg{\varphi(x_0,t_0)((\partial_t^2-v_F^2\partial_x^2)j^1(x,t)+\frac{1}{\pi}\partial_t^2\varphi(x,t))} = 0$$

Plugging this result into our original Schwinger-Dyson equation, we have

$$-i\delta(x-x_0)\delta(t-t_0) =  \biggr(-\frac{1}{g^2}(\partial_{t}^2-v_B^2\partial_{x}^2) - \frac{v_F}{\pi}\frac{\partial_t^2}{\partial_t^2-v_F^2\partial_x^2}\biggr)\bigavg{\varphi(x_0,t_0)\varphi(x,t)}$$

This equation clearly identifies the parenthetical expression on the left-hand side as the inverse propagator for the $\varphi$-field; upon passing to Fourier space, we arrive at Eq. \ref{eq:BosonPropNonPert}.

\section{Bosonization of the 1D dipole conserving model}
\label{Sec:AppBosonization}

We start with the low-energy Dirac fermion action in Eq. \ref{eq:DiracAction} from the main text. Following Ref. \onlinecite{fradkin_2013}, we write $\psi_{R/L}(x)\sim e^{i\sqrt{\pi}(\vartheta\pm\phi)}$. We note that $\vartheta(x)\mapsto \vartheta(x)+cx/\sqrt{\pi}$ under the dipole symmetry transform (with $\varphi\mapsto \varphi+c$). Identifying the current operators $j_0(x)=\psi^\dag(x)\psi(x)$, $j_{1}(x)=\psi^\dag(x)\sigma_z\psi(x)$, and observing that these operators are invariant under the dipole symmetry, we have the operator equivalence 
\begin{equation}\label{eq:BosonizationMappingApp}
j_{0}(x)= (1/\sqrt{\pi})\partial_x\phi,\,\,\,j_{1}(x)= -(1/\sqrt{\pi})(\partial_x\vartheta-\varphi/\sqrt{\pi})
\end{equation}

Accordingly, we may write down a bosonized theory of Eq. \ref{eq:DiracAction}:
\begin{equation}\label{eq:1DLLactionWithDualFieldApp}
    \mathcal{S} = \frac{1}{2}\int dt dx\, \biggr(-v_F(\partial_x\vartheta-\varphi/\sqrt{\pi})^2-v_F(\partial_x\phi)^2
    +2(\partial_t\phi)(\partial_x\vartheta)\biggr)  + \mathcal{S}_{NGB}
\end{equation}
One can check the equations of motion of Eq. \ref{eq:1DLLactionWithDualFieldApp} reproduce the current conservation $\partial_\mu j^\mu = 0$, the anomaly equation $\partial_\mu j^\mu_5=-\partial_t \varphi/\pi$, and the $\varphi_x$ equation of motion $\Box\varphi = -v_Fj_1$. We can integrate out either the $\vartheta$ or $\phi$ fields to get two dual actions:
\begin{equation}\label{eq:1DLLactionAppPhi}
    \mathcal{S}_\phi = \frac{1}{2}\int dt dx\, \biggr(\frac{1}{v_F}(\partial_t\phi-v_F\varphi/\sqrt{\pi})^2-v_F(\partial_x\phi)^2 - \frac{v_F}{\pi}\varphi^2 \biggr) + \mathcal{S}_{NGB}
\end{equation}
\begin{equation}\label{eq:1DLLactionAppTheta}
    \mathcal{S}_\vartheta = \frac{1}{2}\int dt dx\, \biggr(\frac{1}{v_F}(\partial_t\vartheta)^2-v_F(\partial_x\vartheta-\varphi/\sqrt{\pi})^2\biggr) + \mathcal{S}_{NGB}
\end{equation}

As a sanity check, either action (\ref{eq:1DLLactionAppPhi}, \ref{eq:1DLLactionAppTheta}) reproduces the $\varphi$ self-energy computed in Eq. \ref{eq:BosonPropNonPert}. Furthermore, in the action \ref{eq:1DLLactionAppPhi}, we see that the $\varphi$ field couples to the (spacelike part of) the topological current $\epsilon^{\mu\nu}\partial_\nu\phi$ for the $\phi$-field, which is known to correspond to the electromagnetic current of the fermions. Similarly, $\varphi$ couples to the spacelike part of $\partial_\mu\vartheta-\varphi_\mu/\sqrt{\pi}$, which also corresponds to the current in the fermionic picture. 

We may include the usual Luttinger liquid interactions $\hat{H}_{int} =\frac{\pi}{2}( V^\phi j_0(x)^2+V^\theta(j_1(x))^2)$ in this framework. Using the aforementioned mappings \ref{eq:BosonizationMappingApp}, we modify the bosonized actions by changing $v_F\to v$ and scaling the $\phi$ ($\vartheta$) action by $\kappa$ ($\kappa^{-1}$), where $v=\sqrt{(v_F+V^\phi)(v_F+V^\theta)}$ and $\kappa=\sqrt{(v_F+V^\phi)/(v_F+V^\theta)}$ are the usual Luttinger parameters.


It is a simple matter to compute the correlation functions of the $\vartheta$, $\phi$ fields, given the action Eq. \ref{eq:1DLLactionWithDualFieldApp} with arbitrary Luttinger parameters: 
\begin{equation}\label{eq:1DLLactionMostGeneral}
    \mathcal{S} = \frac{1}{2}\int dt dx\, \biggr(-\frac{v}{\kappa}(\partial_x\vartheta-\varphi_x/\sqrt{\pi})^2-v\kappa(\partial_x\phi)^2
    +2(\partial_t\phi)(\partial_x\vartheta)\biggr)  -\frac{1}{2g^2}\int d\omega dk_x\,\varphi_x(\partial_t^2-v_B^2\partial_x^2)\varphi_x
\end{equation}

We compute the inverse propagator
\begin{equation}
    \biggr\langle\begin{pmatrix}\vartheta \\ \phi \end{pmatrix}\begin{pmatrix}\vartheta & \phi \end{pmatrix}\biggr\rangle^{-1} =\begin{pmatrix}-\frac{v}{\kappa}q^2 - \frac{v^2}{\kappa^2}\frac{g^2}{\pi}\frac{q^2}{\omega^2-v_B^2q^2-\frac{v}{\kappa}\frac{g^2}{\pi}} & \omega q \\ \omega q & -v\kappa q^2\end{pmatrix}
\end{equation}
\begin{equation}
    \implies \biggr\langle\begin{pmatrix}\vartheta \\ \phi \end{pmatrix}\begin{pmatrix}\vartheta & \phi \end{pmatrix}\biggr\rangle = \frac{1}{\omega^2-v^2q^2-\frac{v^3g^2}{\kappa \pi}\frac{q^2}{\omega^2-v_B^2q^2-\frac{v}{\kappa}\frac{g^2}{\pi}}} \begin{pmatrix}v\kappa  & \omega/q \\ \omega/q & \frac{v}{\kappa} + \frac{v^2}{\kappa^2}\frac{g^2}{\pi}\frac{1}{\omega^2-v_B^2q^2-\frac{v}{\kappa}\frac{g^2}{\pi}}\end{pmatrix}
\end{equation}

In the long-wavelength limit $\omega^2,v_B^2q^2\ll (v/\kappa)g^2/\pi = m^2$, we approximate the above as
\begin{equation}
    \implies \biggr\langle\begin{pmatrix}\vartheta \\ \phi \end{pmatrix}\begin{pmatrix}\vartheta & \phi \end{pmatrix}\biggr\rangle \approx \frac{1}{\omega^2\biggr(1+\frac{v^2q^2}{m^2}\biggr) - \frac{v^2v_B^2q^4}{m^2}} \begin{pmatrix}v\kappa  & \omega/q \\ \omega/q & -v\frac{\omega^2-v_B^2q^2}{m^2}\end{pmatrix}  
\end{equation}

Using the representation $\psi_{R/L}\sim e^{i\sqrt{\pi}(\vartheta\pm\phi)}$, we identify the fermion correlation function
\begin{equation}
    \avg{\psi_R(x,t)\psi^\dag_R(0)} \sim \exp(\pi(\avg{\vartheta\vartheta}+\avg{\phi\phi}+2\avg{\vartheta\phi}))
\end{equation}
where $\avg{\vartheta\vartheta} = \avg{\mathcal{T}(\vartheta(x,t)\vartheta(x,0))}$ (and likewise for other fields). When the contour integration is performed over the $\omega$-axis in evaluating these Fourier transforms, the factors $\omega/q\sim \abs{q}$ and $\omega^2-q^2\sim q^2$ at low momentum. Thus, the only relevant piece to compute for $x,t\gg m^2$ is $\avg{\vartheta\vartheta}$, which may be further simplified in the long-wavelength limit as
\begin{equation}
    \avg{\mathcal{T}(\vartheta(x,t)\vartheta(0))}  = v\kappa \int\frac{dqd\omega}{(2\pi)^2}e^{i(qx-\omega t)}\frac{1}{\omega^2- \frac{v^2v_B^2q^4}{m^2}}
\end{equation}

It can be shown that $\avg{\psi_R(x,0)\psi^\dag_R(0)}\sim e^{-\abs{x}/\xi}$ and $\avg{\psi_R(0,t)\psi^\dag_R(0)} \sim e^{-(\pi/\sqrt{2})\sqrt{vt/\xi}}$, where $\xi^{-1}=m\kappa/(4\pi v_B)$ defines a confinement lengthscale. Results for $\psi_L$ correlators are analogous. We may also compute the one-particle local density of states for the fermions,
\begin{equation}\label{eq:LDOSconfined}
\mathcal{A}(\omega,x=0)=-\frac{1}{\pi}\mathrm{Im}\biggr(\int_0^\infty e^{i\omega t}(\avg{\psi_R(0,t)\psi^\dag_R(0)}+\avg{\psi_L(0,t)\psi^\dag_L(0)})dt\biggr) \sim \sqrt{\frac{v}{\xi}}\abs{\omega}^{-3/2}\exp(-\frac{\pi^2}{16}\frac{v}{\xi\abs{\omega}}).
\end{equation}

For an ordinary spinless Luttinger liquid without coupling to a dipolar Goldstone, $\avg{\psi_R(x,0)\psi^\dag_R(0)}\sim \abs{x}^{-\frac12(\kappa+\kappa^{-1})}$ and $\avg{\psi_R(0,t)\psi^\dag_R(0)}\sim \abs{t}^{-\frac12(\kappa+\kappa^{-1})}$, leading to a local density of states $\mathcal{A}(\omega,x=0)\sim \abs{\omega}^{\frac12(\kappa+\kappa^{-1})-1}$ \cite{fradkin_2013}. In contrast, the dipole Goldstone mode confines the fermions and turns the power-law behavior of $\mathcal{A}(\omega,x=0)$ into an essential singularity. 

Of course, upon noticing that Eq. \ref{eq:1DLLactionAppTheta} is precisely the 1D field theory of a Bose-Einstein insulator (BEI) studied in Ref.  \onlinecite{LakeDBHM,Lake_Tilted_Chain,zechmann2022fractonic} (with Eq. \ref{eq:1DLLactionAppPhi} given by a simple duality transform), one may have guessed the aforementioned results for fermions by referring to the computations done in these works. For example, the statement that fermions are confined is equivalent to the statement that the single-boson order parameter in a 1D BEI has a short-ranged order instead of the quasi-long-ranged order expected for an ordinary 1+1d superfluid. It has also been recognized in these references that the dipolar symmetry is spontaneously broken with the equal-time correlator $\avg{e^{i\varphi(x)}e^{-i\varphi(0)}}$ exhibiting true LRO; this is a special case of a generalized Mermin-Wagner theorem for multipolar symmetries \cite{MultipolarMerminWagner}.

\section{One-loop Goldstone boson self-energy}
\label{Sec:AppBosonSE}

\begin{figure}[H]
 \centering
\includegraphics[width=0.7\textwidth]{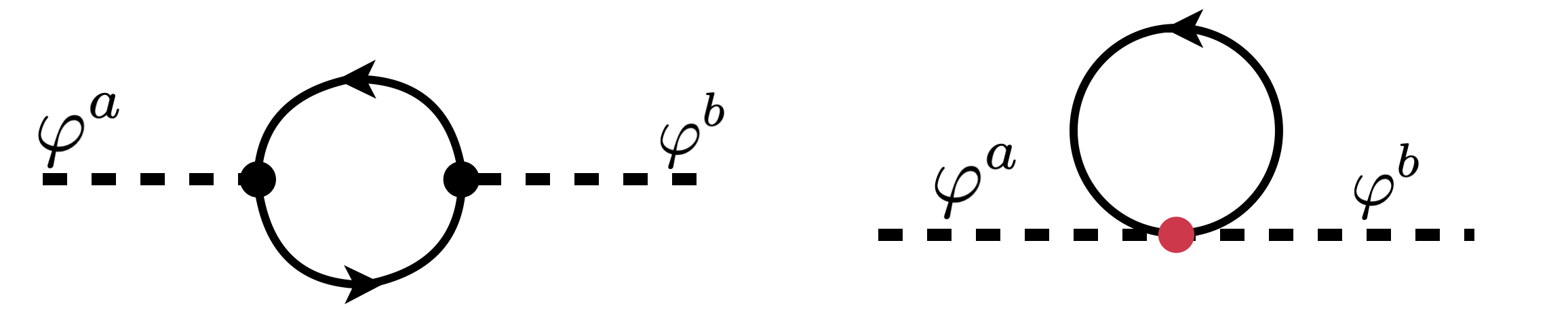}
\caption{1-loop contributions to the Goldstone self-energy.}
 \label{fig:goldstoneSEdiagrams}
\end{figure}

In the low-energy theory of the fermi surface and the NGBs, the 3-point and 4-point interactions give rise to the usual one-loop diagrams for the self-energy (see Fig. \ref{fig:goldstoneSEdiagrams}). The resulting Matsubara sums yield:
\begin{equation}
    \Pi^{ab}(\vec{p},i\omega) = \int \frac{d^dk}{(2\pi)^d} \biggr(\partial_a \xi(\vec{k})\partial_b\xi(\vec{k})\frac{n_F(\xi(\vec{k}+\frac12\vec{p}))-n_F(\xi(\vec{k}-\frac12\vec{p}))}{i\omega-(\xi(\vec{k}+\frac12\vec{p})-\xi(\vec{k}-\frac12\vec{p}))} - n_F(\xi(\vec{k}))\partial_a\partial_b\xi(k)\biggr)
\end{equation}
where $n_F(x)$ is the Fermi-Dirac function. At low momentum $\vec{p}$,
\begin{equation}
    \Pi^{ab}(\vec{p},i\omega) \approx \int \frac{d^dk}{(2\pi)^d} \biggr(\partial_a \xi(\vec{k})\partial_b\xi(\vec{k})\frac{\vec{p}\cdot\grad\xi(\vec{k})}{i\omega-\vec{p}\cdot\grad\xi(\vec{k})}n'_F(\xi(\vec{k})) - n_F(\xi(\vec{k}))\partial_a\partial_b\xi(k)\biggr)
\end{equation}

We may pass the self-energy to a convenient basis $(\vec{a}_1,\vec{a}_2) = (\hat{p}_\parallel, \hat{p}_\perp)$ where $\hat{p}_\parallel=\vec{p}/\abs{\vec{p}}$ and $\hat{p}_\perp$ runs over all the orthogonal directions to $\hat{p}$. Using $\alpha,\beta$ to label components for this new momentum-dependent basis, we re-express the self-energy as
\begin{equation}
    \Pi^{\alpha\beta}(\vec{p},i\omega) \approx \int \frac{d^dk}{(2\pi)^d} \biggr(-\partial_\alpha\epsilon_k\partial_\beta\epsilon_k\frac{\vec{p}\cdot\grad\xi(\vec{k})}{\vec{p}\cdot\grad\xi(\vec{k})-i\omega}n_F'(\xi(\vec{k})) - n_F(\xi(\vec{k}))\partial_\alpha\partial_\beta\epsilon_k\biggr)
\end{equation}

Integrating by parts, and simplifying the resulting expression, 
\begin{equation}
\begin{split}
    \Pi^{\alpha\beta}(\vec{p},i\omega) &=\int \frac{d^dk}{(2\pi)^d} \biggr(-\partial_\alpha\epsilon_k\partial_\beta\epsilon_k\frac{\vec{p}\cdot\grad\xi(\vec{k})}{\vec{p}\cdot\grad\xi(\vec{k})-i\omega}n_F'(\xi(\vec{k})) + \partial_\beta\epsilon_k\partial_\alpha\epsilon_kn_F'(\xi(\vec{k}))\biggr) \\
    &= -i\frac{\omega}{\abs{\vec{p}}}\int \frac{d^dk}{(2\pi)^d} \partial_\alpha\epsilon_k\partial_\beta\epsilon_kn_F'(\xi(\vec{k}))\frac{1}{\partial_\parallel\epsilon_k-i\frac{\omega}{\abs{\vec{p}}}}
\end{split}
\end{equation}

Assuming $\abs{p_0}/\abs{\vec{p}}\ll 1$, we may effectively replace:
\begin{equation}
    \Pi^{\alpha\beta}(\vec{p},i\omega) \approx -i\frac{\omega}{\abs{\vec{p}}}\int \frac{d^dk}{(2\pi)^d} \partial_\alpha\epsilon_k\partial_\beta\epsilon_kn_F'(\xi(\vec{k}))\biggr(\frac{1}{\partial_\parallel \epsilon_k}+i\pi\sign(\omega)\delta(\partial_\parallel \epsilon_k)\biggr)
\end{equation}

Now, we specialize to the 2-D case, and further take the low-temperature limit where $n_F'(\xi)\approx -\delta(\xi)$. Assuming a circular Fermi surface for simplicity, we can estimate $\int \frac{d^2k}{(2\pi)^2}\delta(\xi(\vec{k})) = \frac{1}{(2\pi)^2}\int \frac{1}{v_F}dk_\theta$, where $k_\theta$ parametrizes the Fermi surface. Then, in the $\hat{p}_\perp,\hat{p}_\parallel$ basis, 
\begin{equation}
    \mathbf{\Pi}_{2D}(\vec{p},i\omega) \approx i\frac{\omega}{\abs{\vec{p}}}\frac{1}{(2\pi)^2}\int dk_{\theta}\frac{1}{v_F}\biggr(\begin{pmatrix}  \partial_\parallel\epsilon_k    &  \partial_\perp\epsilon_k \\  \partial_\perp\epsilon_k &  (\partial_\perp\epsilon_k)^2/\partial_\parallel\epsilon_k \end{pmatrix}+i\pi\sign(\omega)\delta(\partial_\parallel\epsilon_k)\begin{pmatrix}  \partial_\parallel\epsilon_k\partial_\parallel\epsilon_k   &  \partial_\parallel\epsilon_k \partial_\perp\epsilon_k\\  \partial_\perp\epsilon_k \partial_\parallel\epsilon_k &  \partial_\perp\epsilon_k\partial_\perp\epsilon_k \end{pmatrix}\biggr)
\end{equation}
Symmetries of the Fermi surface will generally force the first integral to vanish, and the second integral will receive contributions at the antipodal points on the Fermi surface $k_F^\pm$ where $\partial_\parallel\epsilon_k=0$. At the relevant points on the FS, $\partial_\parallel\epsilon_k$ vanishes by construction and $\partial_\perp\epsilon_k=v_F$. Then, adding a factor of $N$ for the fermion flavours, and dividing by a factor of $\kappa$ (the Fermi surface curvature) due to the $\delta(\partial_\parallel\epsilon_k)$, we have
\begin{equation}
    \mathbf{\Pi}_{2D}(\vec{p},i\omega) = -\pi\frac{\kappa^{-1}}{(2\pi)^2}Nv_F\frac{\abs{\omega}}{\abs{\vec{p}}}\begin{pmatrix} 0   &  0\\  0  &  1 \end{pmatrix} = -\frac{Nv_F}{4\pi \kappa}\frac{\abs{\omega}}{\abs{\vec{p}}}(\mathbbm{1}-\hat{p}\hat{p})
\end{equation}
Subsequently, we may write down the one-loop NGB propagator:
\begin{align}\label{eq:NGBPropagator}
D_{ab}(\vec{p},i\omega)&=\frac{g^2}{(\omega^2+v^2_1p^2)\delta_{ab}+v^2_2p_ap_b-g^2\Pi_{2D,ab}(\omega,\vec{p})} \\ \nonumber
&= g^2\frac{\delta_{ab}-p_ap_b/\abs{\vec{p}}^2}{\omega^2+v^2_1\abs{\vec{p}}^2+\gamma g^2\frac{\abs{\omega}}{\abs{\vec{p}}}} + g^2\frac{p_ap_b}{\abs{\vec{p}}^2}\frac{1}{\omega^2+(v^2_1+v^2_2)\abs{\vec{p}}^2}\\
&=D^T_{ab} + D^L_{ab},
\end{align}
where $\gamma=Nv_F/4\pi\kappa$ (note that $v_F$ depends weakly on $\vec{p}$, since it is the Fermi velocity at the spot on the Fermi surface where $\vec{p}$ is tangent). The first term is a Landau-damped pole corresponding to the transverse mode $\varphi_T$, and the second is a standard propagating mode (unchanged from the free theory) corresponding to the longitudinal mode $\varphi_L$. We note the similarity to the computation of a $U(1)$ gauge boson propagator in the literature \cite{pa_lee_gauge_field}, though the longitudinal mode is completely absent in the gauge theory case (as can be made apparent by choosing the Coulomb gauge). 

It is useful to compute $\Pi^{ab}(0,i\omega)$, since this shows up in the expression for the optical conductivity \ref{eq:RPAconductivityCancelling}. Unsurprisingly, considering that $\varphi^a$ couples as a gauge field, we have
\begin{align}\label{eq:SEDrudeWeight}
{\Pi}^{ab}_{2D}(0,i\omega) = \int \frac{d^dk}{(2\pi)^d} \partial_a\epsilon_k\partial_b\epsilon_kn_F'(\xi(\vec{k}))
\end{align}
which is nothing but ($-1$ times) the Drude weight of the Fermi surface (i.e. the coefficient of the $\frac{1}{\omega}$ part of the conductivity tensor for the free fermions, which includes both the current-current correlator and the diamagnetic contributions).
 
In 3D, the computation proceeds similarly. Assuming a spherical Fermi surface for the ease of computation, we suppose that $\vec{p}=\abs{\vec{p}}\hat{z}$, such that the $\delta(\partial_\parallel \epsilon_k)=\delta(\partial_z\epsilon_k)$ restricts the integration to the equator of the Fermi sphere. Passing back to a \textit{Cartesian} basis $a,b\in\{x,y,z\}$, we have
\begin{equation}
    \Pi^{ab}(\vec{p},i\omega) \approx -\sign(\omega)\pi \frac{\omega}{\abs{\vec{p}}}\int \frac{d^3k}{(2\pi)^3} \partial_a\epsilon_k\partial_b\epsilon_k\delta(\xi(\vec{k}))\delta(\partial_z\epsilon_k) = -\sign(\omega)\frac{ k_F}{8\pi^2v_F\kappa}\frac{\omega}{\abs{\vec{p}}}\int_0^{2\pi}d\phi\,\partial_a\epsilon_k\partial_b\epsilon_k.
\end{equation}
Parametrizing $\partial_x\epsilon_k=v_F\cos\phi$, $\partial_y\epsilon_k=v_F\sin\phi$, and $\partial_z\epsilon_k=0$ by construction, we finally have
\begin{align}\label{eq:NGBPropagator3D}
\mathbf{\Pi}_{3D}(\vec{p},i\omega) \approx -\frac{v_F k_F}{16\pi^2\kappa}\frac{\abs{\omega}}{\abs{\vec{p}}}(\mathbbm{1}-\hat{p}\hat{p}).
\end{align}
\section{One-loop fermion self-energy}
\label{Sec:AppFermionSE}

\begin{figure}[H]
 \centering
\includegraphics[width=0.4\textwidth]{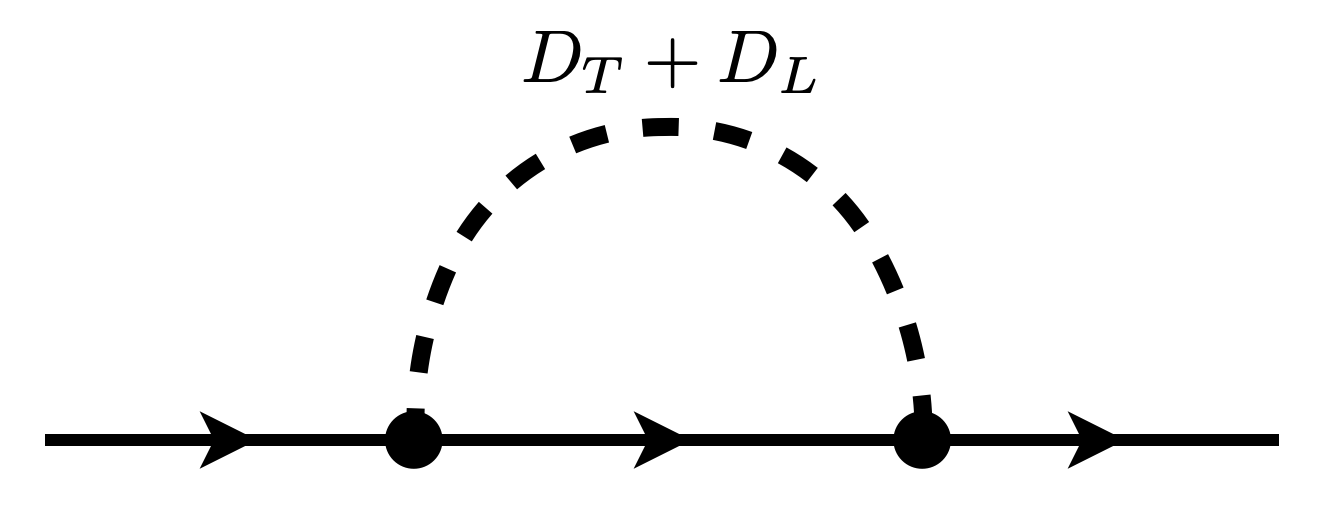}
\caption{1-loop contribution to the fermion self-energy.}
 \label{fig:fermionSEdiagramsI}
\end{figure}

From the diagram in Fig. \ref{fig:fermionSEdiagramsI}, the expression for the fermion self-energy in $(2+1)$-dimensions consists of two parts
\begin{equation}
    \Sigma^{T,L}(i\omega,\vec{q})=g^2\int\frac{d\nu d^2l}{(2\pi)^3}\frac{\partial_a\epsilon(\vec{q}+\frac12\vec{l})\partial_b\epsilon(\vec{q}+\frac12\vec{l})}{i(\omega+\nu)-\epsilon(\vec{q}+\vec{l})}D^{T,L}_{ab}(\nu,\vec{l}).
\end{equation}
We restrict attention to the contribution from $D^T$. The contribution from the longitudinal Goldstone propagator $D_{ab}^L$ will turn out to be subleading to $\Sigma^{T}$ \cite{nayak_wilczek_nFL_fixed_point} in the low-frequency scaling limit that we use; later, we demonstrate this more explicitly. There is another one-loop diagram that one may draw using the 4-point interaction vertex (proportional to $2$ derivatives of the dispersion), but this simply renormalizes the chemical potential. For $\vec{q}$ on the fermi surface, we approximate $\grad\epsilon(\vec{q}+\frac12\vec{l})\approx v_F\hat{w}(\vec{q})$, where $\hat{w}(\vec{q})$ is the normal vector to the fermi surface; we also have $\epsilon(\vec{q}+\vec{l})\approx v_F\hat{w}\cdot\vec{l}+\frac12 \kappa_{ab}l^a l^b$. Thus,
\begin{equation}
    \Sigma^{T}(i\omega,\vec{q})=g^2v_F^2\int\frac{d\nu d^2l}{(2\pi)^3}\frac{\hat{w}_a\hat{w}_b}{i(\omega+\nu)-v_F\hat{w}\cdot\vec{l}-\frac12 \kappa_{ab}l^a l^b}\frac{\delta_{ab}-l_al_b/\abs{\vec{l}}^2}{\nu^2+v^2_1\abs{\vec{l}}^2+\gamma g^2\frac{\abs{\nu}}{\abs{\vec{l}}}}.
\end{equation}
Choosing coordinates such that $\hat{w}=\hat{x}$, we have
\begin{equation}
    \Sigma^{T}(i\omega,\vec{q})=\frac{1}{(2\pi)^3}g^2v_F^2\int d\nu \int dl_x\int dl_y\frac{1}{i(\omega+\nu)-v_Fl_x-\frac12 \kappa_{ab}l^a l^b}\frac{1}{\nu^2+v^2_1(l_x^2+l_y^2)+\gamma g^2{\abs{\nu}}/{\abs{\vec{l}}}}\frac{l_y^2}{l_x^2+l_y^2}
\end{equation}
Following Ref. \onlinecite{sachdevQPT}, we implement a scaling transform $l_x\mapsto b^{-2}l_x'$, $l_y\mapsto b^{-1}l_y'$, $\nu\mapsto b^{-z}\nu'$, and define $\omega=b^{-z}\omega'$. Taking the IR limit $b\to \infty$ corresponds to finding the small-$\omega$ scaling of the self-energy and allows us to discard irrelevant terms in the propagators. In particular, the $\kappa_{ab} l^a l^b$ may be replaced with $\kappa l_y^2$, ${l_y^2}/{(l_x^2+l_y^2)}\to 1$, ${\abs{\nu}}/{\abs{\vec{l}}}\to \abs{\nu/l_y}$, and the terms $\nu^2$, $v^2_1l_x^2$ may be neglected in the denominator of the boson propagator. Note that $\gamma$ depends on $v_F$, which is a function of $\vec{l}$. However, provided the fermi velocity does not vanish anywhere on the fermi surface, we may replace $v_F(\vec{l})$ with $v_F(\vec{q})$, at the cost of terms which will anyway disappear in the scaling limit. Accordingly,
\begin{equation}\label{eq:FermionSelfEnergyIntegralAnalyticallySoluble}
\Sigma^{T}(i\omega,\vec{q})=\frac{1}{(2\pi)^3}g^2v_F^2b^{-(3+z)}\int d\nu' \int dl'_x\int dl'_y\frac{1}{ib^{-z}(\omega'+\nu')-v_Fb^{-2}l'_x-\frac12 \kappa b^{-2}(l_y')^2}\frac{1}{v^2_1b^{-2}(l'_y)^2+\gamma g^2b^{1-z}\abs{\nu'/l'_y}}
\end{equation}
Now, this integral (which is of the same form as the self-energy integrals obtained in the patch decomposition, e.g. in Ref. \onlinecite{metlitski_sachdev_ising}) can be evaluated analytically. We find
\begin{equation}\label{eq:FermionSelfEnergyAppendixMassless}
\Sigma^{T}(i\omega,\vec{q}) = -i\sign(\omega)\biggr(\frac{1}{6\sqrt{3}\pi^2}\frac{g^4v_F^2\kappa}{Nv_1^4}\biggr)^{1/3}\abs{\omega}^{2/3}=-i\sign(\omega)E_{NFL}^{1/3}\abs{\omega}^{2/3}
\end{equation}
which is valid for $\omega\ll E_{NFL}$. This form of the self-energy implies $z=3$ at this level of treatment, analogous to the usual models of metallic quantum criticality in $(2+1)$-dimensions. 

\subsection{Self-Energy from the Longitudinal Goldstone Propagator}

It is useful to estimate the contribution of the longitudinal Goldstone degree of freedom and compare it to the self-energy in Eq. \ref{eq:FermionSelfEnergyAppendixMassless}. To do this, we will perform the computation in the patch theory of Eq. \ref{eq:PatchLagrangianMainText}, having included the irrelevant terms in the Goldstone kinetic term as well as the interaction with the longitudinal Goldstone; we reproduce the fermionic part of the action below.

$$\mathcal{S}_{\theta} =\sum_s\int d\tau d^2x\, \bar{\psi}_{\theta,s}\biggr( \partial_\tau + s_\theta v_F(i\partial_x +\varphi_T) - \frac12\kappa (\partial_y-i\varphi_L)^2\biggr)\psi_{\theta,s}$$

The only vertex needed to extract an $\omega$-dependent self-energy for the fermions is the three-point vertex involving $\varphi_L$, and recall that the RPA propagator of the longitudinal mode contains no Landau-damping term. Restricting to the Fermi surface (i.e. $\vec{q}=0$ in the patch theory), the relevant Feynman diagram thus yields

$$\Sigma^L(i\omega) = -\frac{g^2\kappa^2}{4v_L^2}\int\frac{d\nu dl_ydl_x}{(2\pi)^3}\frac{l_y^2}{i(\omega+\nu)-v_Fl_x-\frac12\kappa l_y^2}\frac{1}{\nu^2/v_L^2+l_y^2+l_x^2}$$

Performing the contour integral over $l_x$ and shifting the $\nu$ variable as needed, we have

$$\Sigma^L(i\omega) = \frac{g^2\kappa^2}{4v_L^2v_F}\frac{2\pi i}{(2\pi)^3} \int d\nu dl_y  l_y^2\biggr[\frac{\vartheta(\nu)}{l_y^2+(\nu-\omega)^2/v_L^2+(\frac12\kappa l_y^2-i\nu)^2/v_F^2} - \frac12\frac{1}{l_y^2+\nu^2/v_L^2-iv_F^{-1}\sqrt{l_y^2+\nu^2/v_L^2}(\frac12\kappa l_y^2-i(\omega+\nu))}\biggr]$$

It is difficult to evaluate this expression exactly, but we may nevertheless show that $\Sigma^L(i\omega)\sim i\omega + \ldots$ where $\ldots$ are higher-order in $\omega$ by computing $\pdv{\Sigma^L}{\omega}$ and checking that it is finite at $\omega=0$ (with a UV cutoff implemented, if need be). Accordingly,

$$\pdv{\Sigma^L}{\omega}\biggr\lvert_{\omega=0} = \frac{g^2\kappa^2}{4v_L^2v_F}\frac{2\pi i}{(2\pi)^3} \int d\nu dl_y  \biggr[\frac{\vartheta(\nu) 2l_y^2\nu/v_L^2}{(l_y^2+\nu^2/v_L^2+(\frac12\kappa l_y^2-i\nu)^2/v_F^2)^2} - \frac{1}{2v_F}\frac{l_y^2\sqrt{l_y^2+\nu^2/v_L^2}}{\biggr(l_y^2+\nu^2/v_L^2-iv_F^{-1}\sqrt{l_y^2+\nu^2/v_L^2}(\frac12\kappa l_y^2-i\nu)\biggr)^2}\biggr]$$

One may check that the above integrals are finite so long as we include a UV cutoff (the necessity of such a cutoff is an artifact of the patch description of the Fermi surface). Thus, we conclude that the leading contribution to the self-energy from the longitudinal Goldstones scales as $\Sigma^L(i\omega)\sim c\times i\omega$, which is subleading to the $\omega^{2/3}$ contribution in Eq. \ref{eq:FermionSelfEnergyAppendixMassless}. Nondimensionalizing the above integral, we identify the constant $c = g^2\kappa^2/v_Fv_L^2 \times F$, where the function $F$ only depends on dimensionless numbers, including the ratios $v_L/v_F$ and $\kappa/v_F$. Supposing that the only contributions to the self-energy are due to the transverse and longitudinal Goldstone bosons, we identify another crossover frequency $\omega^*\sim v_L^2/g^2$ (having neglected $O(1)$ numbers and dimensionless ratios) above which the self-energy from the longitudinal Goldstone exceeds the contribution from the transverse Goldstone. 

\subsection{Pseudo-Goldstone modes}
It is useful to consider the case of adding a finite boson mass $\frac12 r\varphi_T^2$ to the propagator $D^T$. In this case, the small-$\omega$ self-energy is modified to

\begin{equation}\label{eq:FermionSelfEnergyAppendixMassiveExact}
\begin{split}
&\Sigma^T(i\omega,\vec{q}) = -i\sign(\omega)E_{NFL}^{1/3}\abs{\omega}^{2/3}F(\zeta)\\
&F(\zeta) = \frac{\sqrt{3}}{\pi}\int_0^\infty x\log(1+\frac{1}{x^3+\zeta x})dx,\,\,\,\,\zeta = \frac{r}{(v^2_1 g^4\gamma^2)^{1/3}}\omega^{-2/3} = (\omega^*/\omega)^{2/3},
\end{split}
\end{equation}
where the function $F(0)=1$ (i.e. when the boson mass is zero). For nonzero $r$, we may estimate the asymptotic dependence on the dimensionless parameter $\zeta$ by writing
\begin{equation}
    F(\zeta) \approx \frac{\sqrt{3}}{\pi}\biggr(\int_0^{\zeta^{1/2}}x\log(1+\frac{1}{\zeta x})dx + \int_{\zeta^{1/2}}^\infty x\log(1+\frac{1}{x^3})dx\biggr)
\end{equation}
For $\zeta \gg 1$, the first integral dominates and is given $\frac{\sqrt{3}}{\pi}\zeta^{-1/2}$; for $\zeta\ll 1$, the second integral dominates and is equal to unity (up to subleading terms in both cases). As a result, we obtain the scaling form of the self-energy away from the crossover frequency:
\begin{equation}\label{eq:MassiveGoldstoneFermiSEScaling}
 \abs{\Sigma^T(i\omega)}\approx
  \begin{cases}
                                   \frac{\sqrt{3}}{\pi}(E_{NFL}/\omega^*)^{1/3}\omega & \text{\,\,$\omega\ll \omega^*$} \\
                                   E_{NFL}^{1/3}\omega^{2/3} & \text{\,\,$\omega^*\ll\omega\ll E_{NFL}$} 
  \end{cases}
\end{equation}
with the crossover scale $\omega^* = (v_1g^2\gamma)^{-1}r^{3/2}$ defined by $\zeta=1$; the coefficient $\sqrt{3}/\pi$ in front of the small-$\omega$ self-energy deviates only slightly from numerical evaluation of $F(\zeta)$. Replacing $\omega$ with the Matsubara frequency $\pi T$, this suggests a crossover temperature $T^* \sim r^{3/2}$, above which the system behaves like a non-Fermi liquid and below which the system behaves like a fermi liquid. This computation is very similar to the one performed in Ref. \onlinecite{mandal_boson_mass_NFL_crossover}, where the NFL/FL crossover frequency of a fermi surface coupled to pseudo-Goldstone bosons is also predicted to scale like $r^{3/2}$. 

\subsection{Interaction with a single Goldstone Boson}

Supposing that the system only conserves the $x$-component of the dipole moment (thus only $\varphi^x$ is a gapless mode that can couple strongly to the Fermi surface), then it follows that the boson propagator (using the most general free action consistent with cubic symmetry) is 
\begin{equation}
    D(i\omega,\vec{p})=\avg{\varphi_x\varphi_x} = \frac{1}{\omega^2+v^2_1\abs{p}^2+v^2_2p_x^2 + \gamma g^2\abs{\omega/\vec{p}}\sin^2\theta_p}
\end{equation}
where $\theta_p=\arctan(p_y/p_x)$. This self-energy can be found simply by restricting Eq. \ref{eq:OneLoopSelfEnergy} to the $\hat{x}\hat{x}$-component. Using this to compute the fermion self-energy with a diagram analogous to Fig. \ref{fig:fermionSEdiagramsI}, we have
\begin{equation}
    \Sigma(i\omega,\vec{q}) = g^2\int \frac{d\nu d^2l}{(2\pi)^3} \frac{\partial_x\epsilon(\vec{q}+\frac12\vec{l})\partial_x\epsilon(\vec{q}+\frac12\vec{l})}{i(\omega+\nu)-\epsilon(\vec{q}+\vec{l})}\frac{1}{\nu^2+v^2_1\abs{l}^2+v^2_2l_x^2 + \gamma g^2\abs{\nu/\vec{l}}\sin^2\theta_l}.
\end{equation}
Again restricting to $\vec{q}$ on the fermi surface and rotating coordinates to $\hat{w}(\vec{q}) = \hat{x}$ (where $\hat{w}(\vec{q})$ is the normal vector to the fermi surface), we have
\begin{equation}
    \Sigma(i\omega,\vec{q}) = g^2v_F^2\int \frac{d\nu d^2l}{(2\pi)^3} \frac{\cos^2\theta_F}{i(\omega+\nu)-v_Fl_x-\frac12\kappa_{ab}l^al^b}\frac{1}{\nu^2+v^2_1\abs{l}^2+v^2_2l^2\cos^2(\theta_l-\theta_F) + \gamma g^2\abs{\nu/\vec{l}}\sin^2(\theta_l-\theta_F)}
\end{equation}
where $\theta_F$ is the angle between $\hat{w}(\vec{q})$ and the $x$-axis. Applying the same scaling transformation as before to extract the small-$\omega$ behavior of $\Sigma(\omega,\vec{q})$, we have (after discarding terms that are subleading in the scaling limit):
\begin{equation}
    \Sigma(i\omega,\vec{q}) = g^2v_F^2b^{-(3+z)}\int \frac{d\nu' dl_y'l_x'}{(2\pi)^3} \frac{\cos^2\theta_F}{ib^{-z}(\omega'+\nu')-v_Fb^{-2}l'_x-\frac12\kappa b^{-2}(l_y')^2}\frac{1}{(v^2_1+v^2_2\sin^2\theta_F)b^{-2}(l_y')^2 + \gamma g^2b^{1-z}\abs{\nu'/l_y'}\cos^2(\theta_F)}
\end{equation}
This is an integral with the same structure as in Eq. \ref{eq:FermionSelfEnergyIntegralAnalyticallySoluble}, with the only difference being that $g^2\to g^2\cos^2\theta_F$ and $v^2_1\to v^2_1+v^2_2\sin^2\theta_F$. Hence the fermion self-energy is given by
\begin{equation}\label{eq:FermionSelfEnergyAppendixAnisotropic}
\Sigma(i\omega,\vec{q}) = -i\sign(\omega)\biggr(\frac{1}{6\sqrt{3}\pi^2}\frac{g^4v_F^2\kappa\cos^4\theta_F}{N(v^2_1+v^2_2\sin^2\theta_F)^2}\biggr)^{1/3}\abs{\omega}^{2/3}=-i\sign(\omega)E_{NFL}^{1/3}(\vec{q})\abs{\omega}^{2/3}.
\end{equation}

\section{IR patch action and mapping to the U(1) gauge model}
\label{Sec:AppPatchTheory}

\begin{figure}
    \centering
    \includegraphics[width=0.4\linewidth]{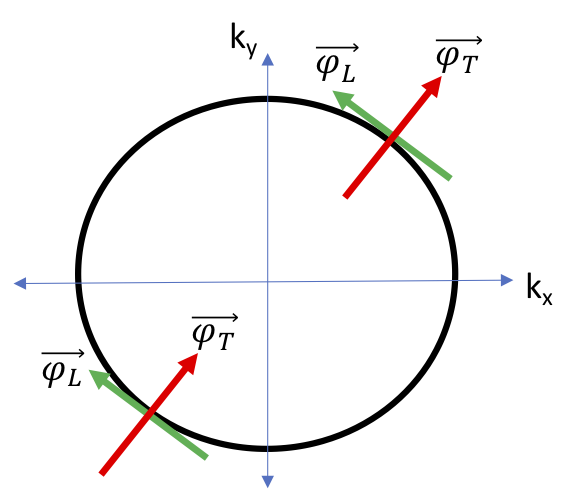}
    \caption{Schematic depiction of the couplings of $\varphi_L$, $\varphi_T$ to the Fermi surface; energetic constraints that force $\varphi_T$ ($\varphi_L$) to be perpendicular (parallel) to the relevant Fermi surface patches, hence the interaction of $\varphi_L$ with the fermions via the covariant derivative is always irrelevant.}
    \label{fig:transverseNGBcoupling}
\end{figure}

Starting with the 2D theory in Eq. \ref{eq:MicroscopicAction}, we write a patch action below, i.e. the two antipodal patches along with the Goldstone modes with momentum $\vec{p}$ tangent to the patches, with $\theta$ serving as the patch index. We rotate our coordinates so that the $x$-direction denotes the direction perpendicular to the fermi surface patches, and the $y$-direction denotes the parallel direction. Recall that the Fermion-Goldstone coupling is exactly that of a $U(1)$ gauge field, with $(0,\varphi^x,\varphi^y)$ playing the role of the vector potential. As a result, 
\begin{equation}
    \begin{split}
        \mathcal{S} &= \mathcal{S}_{GB}[\varphi_x,\varphi_y] + \sum_{\theta\in\pm}\mathcal{S}_{\theta}\\
        \mathcal{S}_{\theta} &=\int dtdxdy\, \bar{\psi}_\theta\biggr(\partial_t + is_\theta v_FD_x - \frac12\kappa_{ab}D_aD_b\biggr)\psi_\theta
    \end{split}
\end{equation}
Here, we have expanded the fermion dispersion $\xi(k) = v_Fk_x+\frac12\kappa_{ab}k_ak_b$ near the patches and defined the covariant derivative $D_a = \partial_a - i\phi_a(t,\vec{x})$. We define $s_\theta$ to be $+1$ ($-1$) on the $+$ ($-$) patch. Since the momentum of the Goldstone boson is in the $\hat{y}$-direction per our choice of coordinates, we see that $\varphi_x$ ($\varphi_y$) plays the role of the $\varphi_T$ ($\varphi_L$) in this patch picture, and thus we expect that couplings via the covariant derivative $D_y$ are irrelevant since $\varphi_L$ is not Landau-damped by the interaction with fermions. This situation is illustrated in Fig. \ref{fig:transverseNGBcoupling}.

Indeed, one may recapitulate computations of the boson and fermion self-energies using this two-patch action, following e.g. Ref. \onlinecite{sslee2009,metlitski_sachdev_ising,mross_mcgreevy_liu_senthil_NFL_eps_N}, and find that the propagator of $\varphi_x$ receives a Landau-damping self-energy correction $\sim \gamma\abs{\omega/q_y}$, while $\varphi_y$ does not, and the fermion self-energy is proportional to $i\sign(\omega)\abs{\omega}^{2/3}$. To get the simplest low energy theory, we implement a scaling transform $\omega\mapsto b^{-z}\omega'$, $q_x\mapsto b^{-2}q_x'$, $q_y\mapsto b^{-1}q_y'$, and scale the boson and fermion fields to keep the leading terms of the one-loop effective action invariant. Accordingly, we choose $\psi_\theta(\omega,\vec{k})\to b^4\psi'_\theta(\omega',\vec{k}')$, $\varphi_T(\omega,\vec{k})\to b^4\varphi_T(\omega',\vec{k}')$ and $z=3$. Applying this to the tree-level patch action and dropping (almost) all irrelevant terms in this scaling limit\footnote{We have not dropped the term $\bar{\psi}_\theta \partial_t \psi_\theta$ even though it is irrelevant under $z=3$ scaling, because doing so would kill any dynamics in the action \ref{eq:FinalPatchActionApp}. Including this irrelevant term ensures that the frequency contour integrals in the one-loop diagrams reproduce the correct results \cite{sslee2009}.}, we are left with
\begin{equation}\label{eq:FinalPatchActionApp}
\begin{split}
    \mathcal{S}_{GB} &= \int \frac{d\omega d^2q}{(2\pi)^3}\frac{1}{2g^2}\biggr(v_1^2q_y^2\varphi^2_x + (v_1^2+v_2^2)q_y^2\varphi^2_y\biggr)\\
    \mathcal{S}_{\theta} &=\int dtdxdy\, \bar{\psi}_\theta\biggr(\partial_t + iv_FD_x - \frac12\kappa \partial_y^2\biggr)\psi_\theta
\end{split}
\end{equation}

with $\kappa=\kappa_{yy}$. We see that the $\varphi_y$ field decouples from the fermi surface in the IR as expected, and the remainder of the theory precisely coincides with an analogous IR patch theory of a fermi surface coupled to a dynamical $U(1)$ gauge field in the Coulomb gauge. The field $\varphi_x$ plays the role of the single physical mode in the gauge theory, but the field $\varphi_y$ has no analog in the gauge theory (as a longitudinal mode is pure gauge and is not a propagating mode). However, since $\varphi_y$ anyway decouples from the fermi surface in the IR, the mapping to the gauge model is still robust as far as universal properties are concerned.

\section{Effects of Cubic Anisotropy}
\label{Sec:AppAnisotropy}

We now recapitulate some of the prior analysis in the case where $v_3^2\neq 0$ in Eq. \ref{eq:MicroscopicAction}. For simplicity, we will restrict to 2 spatial dimensions, though it is straightforward to generalize to higher dimensions. Such a term due to cubic anisotropy will serve to mix the transverse and longitudinal Goldstone modes (along with providing a slight anisotropic renormalization to their individual dispersions). Writing the real-time inverse propagator for the Goldstones in the $\varphi_T,\varphi_L$ basis, we have

$$\mathbf{D}^{-1} = \omega^2\mathbbm{1} - p^2\begin{pmatrix}v_1^2+i\gamma g^2\frac{{\omega}}{p^3}+v_3^2\cos^2\theta\sin^2\theta & \frac12v_3^2\sin(4\theta)\\\frac12v_3^2\sin(4\theta) & v_1^2+v_2^2+v_3^2(\sin^4\theta+\cos^4\theta)\end{pmatrix}$$

where $\vec{p} = (p\cos\theta,p\sin\theta)$ and $\gamma$ is as computed in Sec. \ref{Sec:AppBosonSE}. In the limit $v_3^2\ll \gamma g^2\omega/p^3\sim v_1^2/\gamma g^2$, the poles will be given by:

$$\omega^2 - i\gamma g^2\frac{\omega}{p} -\biggr(v_1^2 + \frac12v_3^2\biggr(1-\cos^2\theta\sin^2\theta+\frac{1-3\cos^2\theta\sin^2\theta}{i\gamma g^2\frac{\omega}{p^3}-v_2^2}\biggr)\biggr)p^2=0$$
$$\omega^2  -\biggr(v_1^2+v_2^2 + \frac12v_3^2\biggr(1-\cos^2\theta\sin^2\theta-\frac{1-3\cos^2\theta\sin^2\theta}{i\gamma g^2\frac{\omega}{p^3}-v_2^2}\biggr)\biggr)p^2=0$$

For $v_3^2=0$, these reproduce the dispersions for $\varphi_T$ and $\varphi_L$ respectively. For small values of $v_3^2$, we can use a self-consistency argument to see that the pole in the first line is still a Landau-damping pole $\omega\sim ip^3$ with coefficient slightly modified to have some angular dependence. Likewise, the second pole is still linearly dispersing, with the only new effect being that the dispersion also has an imaginary part with a leading $p^2$ dependence.

Defining longitudinal and transverse projectors respectively as $\mathcal{P}_{ab}=p_ap_b/p^2$ and $\mathcal{T}_{ab} = \delta_{ab}-\mathcal{P}_{ab}$, first-order perturbation theory in the parameter $v_3^2$ gives the imaginary-time Goldstone propagator as

\begin{equation}\label{eq:NGBPropagatorAppAnisotropic}
\begin{split}
\frac{1}{g^2}D_{ab}(i\omega,\vec{p})&= \frac{\mathcal{T}_{ab} - v_3^2F(\theta)\mathcal{P}_{ab}}{\omega^2+v_T^2(\theta)\abs{\vec{p}}^2+\gamma g^2\frac{\abs{\omega}}{\abs{\vec{p}}}} + \frac{\mathcal{P}_{ab}+v_3^2F(\theta)\mathcal{T}_{ab}}{\omega^2+v_L^2(\theta)\abs{\vec{p}}^2}\\
&=\tilde{D}^T_{ab} + \tilde{D}^L_{ab}\\
v_T^2(\theta)&=v_1^2+\frac12v_3^2\biggr(1-\cos^2\theta\sin^2\theta-\frac{1-3\cos^2\theta\sin^2\theta}{v_1^2+v_2^2}\biggr)\biggr) + O(v_3^4)\\
v_L^2(\theta)&=v_1^2+v_2^2+\frac12v_3^2\biggr(1-\cos^2\theta\sin^2\theta\biggr) + O(v_3^4)\\
F(\theta) &= \frac{\frac12\sin(4\theta)}{v_2^2+v_3^2(\sin^4\theta+\cos^4\theta-\cos^2\theta\sin^2\theta)-\gamma g^2\omega/p^3}
\end{split}
\end{equation}

Since the anisotropy term mixes $\varphi_L$ and $\varphi_T$, the part of the Goldstone propagator with the Landau-damping pole (which drives the non-Fermi Liquid behavior of the system) will acquire a contribution proportional to the longitudinal projector $\mathcal{P}_{ab}=p_ap_b/p^2$, with the coefficient of the transverse projector $\mathcal{T}_{ab} = \delta_{ab}-\mathcal{P}_{ab}$ unchanged at $O(v_3^2)$. Thus, due to the anisotropy term, the strongly-coupled Goldstone mode is no longer purely transverse. Nevertheless, when performing the fermion self-energy integrals in Sec. \ref{Sec:AppFermionSE} (Fig. \ref{fig:fermionSEdiagramsI}) with the Landau-damped part of the propagator (which is the only integral that contributes to a $\omega^{2/3}$ self-energy), only the contribution due to the part of $\tilde{D}^T_{ab}$ proportional to $\mathcal{T}_{ab}$ will contribute to a $\omega^{2/3}$ self-energy, with the other three integrals giving rise to subleading corrections. Furthermore, the scaling limit discussed in \ref{Sec:AppFermionSE} allows us to take $\theta=\pi/2$ as it appears in the angular dependence of $v_T^2$.  

Accordingly, the only notable differences from the isotropic case are an additional small patch-angle ($\theta_F$) dependence in the energy scale $E_{NFL}$ (relative to the formula Eq. \ref{eq:FermionSelfEnergyAppendixMassless}, this comes from taking $v_1^2\to v_T^2(\pi/2 - \theta_{F})$) and a small damping for the mode that decouples from the Fermi surface in the IR (the latter of which is anyway invisible in the IR limit of the patch decomposition). This new patch-angle dependence of the boson dispersion is not present for a Fermi surface coupled to a $U(1)$ gauge field, but such a dependence is anyway washed out in the deep IR, since a patch-dependent renormalization to the boson dispersion is absorbed into the fermion-boson coupling $g^2$ at each patch, which all flow to the same fixed-point value $g^2_*$. Hence, we conclude that the proposed mapping to the IR physics of the $U(1)$ gauge model goes through even with an anisotropic dispersion in the Goldstone kinetic term. 

A quick way to understand this result is by referring to the arguments of the previous Sec. \ref{Sec:AppPatchTheory}. The coupling between $\varphi_L$ and the fermions is always irrelevant, so even if cubic anisotropy can scatter $\varphi_L$ and $\varphi_T$ into each other, the only relevant effect is to renormalize the coefficient $v_1^2$ at each pair of patches since $\varphi_L$ does not couple to the fermions in this IR theory.

\end{document}